\documentclass[final,3p,times]{elsarticle}
\usepackage{graphicx}
\usepackage{color}
\usepackage{dcolumn}
\usepackage{bm}
\usepackage{natbib}
\usepackage{notoccite}
\usepackage{graphics}
\usepackage{amssymb}
\usepackage{times}
\usepackage{subfig}
\usepackage{amsbsy}
\usepackage{amsmath}
\usepackage{amsfonts}
\usepackage{amsthm}
\usepackage{float}
\usepackage{textcomp}
\DeclareMathOperator{\tr}{Tr}
\newcommand{\ket}[1]{| #1 \rangle}
\newcommand{\bra}[1]{\langle #1 |}
\newcommand{\braket}[2]{\langle #1 | #2 \rangle}
\newcommand{\ketbra}[2]{| #1 \rangle \langle #2 |}

\pdfoutput=1
\theoremstyle{plain}

\theoremstyle{definition}

\begin{document}

\title{Quantum mutual information and quantumness vectors for multiqubit systems}
\author{Sk Sazim}
\ead{sk.sazimsq49@gmail.com}
\author{Pankaj Agrawal}
\ead{agrawal@iopb.res.in}
\address{Institute of Physics, Sainik School Post,
Bhubaneswar-751005, Orissa, India.}

\begin{abstract}
We introduce a new information theoretic measure of quantum correlations for multiparticle 
systems. We use a form of multivariate mutual information -- the interaction information 
and generalize it to multiparticle quantum systems. There are a number of different possible 
generalizations. We consider two of them. One of them is related to the notion of 
quantum discord and the other to the concept of quantum dissension. 
This new measure, called dissension vector, is a set of numbers -- quantumness vector. 
This can be thought of as a fine-grained measure, as opposed to measures that
quantify some average quantum properties of a system. These 
quantities quantify/characterize the correlations present in multiparticle states. 
We consider some multiqubit states and find that these quantities are responsive to different  
aspects of quantumness, and correlations present in a state. We find that different dissension vectors 
can track the correlations (both classical and quantum), or quantumness  only.
As physical applications, we find that these vectors might be useful in several information 
processing tasks. We consider the role of dissension vectors -- (a) in deciding the 
security of BB84 protocol against an 
eavesdropper and (b) in determining the possible role of correlations in the performance of Grover 
search algorithm. Specially, in the Grover search algorithm, we find that dissension
vectors can detect the correlations and show the maximum correlations when one expects.

\end{abstract}

\maketitle
\section{Introduction}
In the quantum information science, one of the challenges is to understand the nature of
 correlations present in a multiparticle system. Because of its complex nature, we still have
little success in this respect \cite{RevModPhys.84.1655, RevModPhys.81.865}. Correlations, specially quantum
correlations, have been very useful for a host of quantum information processing tasks,
such as quantum computing \cite{2005quant.ph..8124J, PhysRevA.84.022324, PhysRevLett.100.050502}, 
quantum cryptography \cite{RevModPhys.74.145}, quantum metrology \cite{PhysRevLett.96.010401}.
The quantum correlations also lie at the heart of quantum mysteries and account
for many counter-intuitive features of the quantum world. Therefore, a
understanding of the nature of quantum correlations is very important.

The correlations in a system can be of classical and/or quantum nature. Usually,
it is believed that quantum correlations are due to entanglement \cite{RevModPhys.81.865}.
However, more recently, it has been suggested that the quantum correlations go
beyond the simple idea of entanglement \cite{RevModPhys.84.1655}. In particular,
it has been argued that quantum
discord \cite{PhysRevLett.88.017901, discordHend2001} quantifies all types of quantum correlations
including entanglement. Discord is the difference between total correlations and
classical correlations present in a state. In recent years, it has been one of the
main topics of research \cite{RevModPhys.84.1655, PhysRevA.71.062307, PhysRevLett.100.140502, 
PhysRevLett.104.080501, okrasa2011, PhysRevA.84.042109, Chakrabarty2011}. It has been shown that quantum
discord also has operational significance \cite{PhysRevLett.100.050502, lafla2002, PhysRevA.83.032323}. 
A number of different measures of quantum correlations similar to quantum discord have been 
proposed in the literature. These are quantum deficit \cite{PhysRevLett.90.100402, PhysRevLett.89.180402}, 
measurement induced disturbance \cite{PhysRevA.66.022104}, geometric discord \cite{PhysRevLett.105.190502}, 
and many more.

It has recently been argued that the quantum discord, and similar other 
information theoretic measures, actually not only quantifies entanglement, i.e., nonlocal
quantumness but also local quantumness  \cite{2015arXiv150200857A, PhysRevA.93.062322}. For example, 
due to the presence of local quantumness, the quantum discord (and other related measures \cite{RevModPhys.84.1655}) 
increases by 
applying certain kinds of local noise \cite{PhysRevLett.107.170502}.
It is still to be established that a state with zero entanglement and non-zero
discord can act as a resource for a nonlocal task (cf., \cite{PhysRevA.88.022315, PhysRevLett.112.140507}). 
In this sense, the phrase
``quantum correlations beyond entanglement'' may be a misnomer. However information
theoretic measures like discord do seem to characterize quantum properties of a
state beyond entanglement, in particular local quantumness. Such measures
appear to characterize the quantum properties of a state more completely. Therefore,
it will be useful to generalize the measures like quantum discord to multiparticle systems.
There have been several attempts in this direction 
\cite{RevModPhys.84.1655, PhysRevLett.104.080501, PhysRevLett.107.190501, Chakrabarty2011}. 
We will use multivariate mutual information for our generalization.

There exist a number of different versions of multivariate mutual information \cite{cover1991}. 
We consider three different versions -- interaction information, total information, and 
binding information. The extension of the definitions of these versions to quantum world
present myriad possibilities. However, not all the generalizations seem to have
clear physical meaning. We will particularly focus on the generalization of interaction
information to the quantum regime. This is because, classically, interaction information
corresponds to genuine multivariate correlations.

One important point that we emphasize in this paper is the usefulness of a vector-like
quantity to characterize and quantify the quantumness of a state. The correlations
in mixed states of a system, or even pure states of a multiparticle system are multifaceted.
They cannot be characterized by just one number. The set of numbers is called, more
generally, a quantumness vector. We can think of this as a fine-grained measure. A suitable
one number can quantify some average quantum properties of a state which may be suitable
for some applications. We first illustrate it by 
considering two-qubit mixed states. We introduce a quantumness vector for 
characterizing these mixed states. This idea is then extended to multiparticle
states. For generalization of quantum discord to $n$-qubit case, we use interaction
information, a version of multivariate
mutual information \cite{Chakrabarty2011} that characterizes
genuine multivariate correlations in $n$ random variables. It is based on
a Venn-diagram type approach. There exist many expressions for this 
$n$-variable mutual information, all of which are same
classically but differ when conditional entropies are generalized to quantum level. 
For a multiparticle system, one can make measurement on one-particle, or on more 
than one-particle to probe the different aspects of quantum correlations. This
would lead to multiple quantities that can eventually characterize the correlations 
present in the system.  Such physical quantities, quantum dissension, 
were introduced in our previous work \cite{Chakrabarty2011}. By emphasizing that, we
need a set of numbers to characterize correlations, we extend the notion of 
dissension to dissension vector along two different tracks. In the first track
we proceed in the usual way by which quantum discord was defined as difference of 
classical information from total amount of information present in the system. 
Then we extend the definition to multiparticle case. In the second approach, we consider all possible
measurements in the expression of mutual information. In each track, 
to characterize multiparticle correlations, we will have 
$n-1$ quantities based on $(n-1)$ types of measurements.
For example, in the tripartite case, in each track we shall have two
quantities that will characterize the correlations. Interestingly, these
values can be negative because a measurement on a subsystem can enhance the correlations in the 
rest of the system. In the case of three-qubit systems, we find that dissension
vectors, $\vec{\delta}_1^1$, $\vec{\delta}_1^2$ in track-I and in track-II, 
which are based on one-particle measurement,  quantify correlations, both classical and
quantum; more correlations leads to a more negative value. On the other hand, the
dissension vector $ \vec{\delta}_2^1$ which is based on two-qubit measurement
quantifies quantumness of the state - both local and nonlocal.
This approach emphasizes the fact that a single 
quantity alone is not sufficient to characterize the quantum properties of a state.
This paves the way for defining quantum correlation as a vector quantity. 

We find that these dissension vectors are useful in capturing, local as well as 
nonlocal quantumness of the multiparticle states. They reveal the complex structures of correlations present in the state. We also find that 
under nonunital channel, these measures can increase. This suggests that a dissension vector
is also characterizing local quantum properties. To put these measures on strong footing, 
we consider some physical applications. We posit two such applications in quantum 
informations protocols -- a) Bennett and Brassard quantum key distribution protocol 
(BB84) \cite{QKD_BB84} and b) Grover search algorithm \cite{PhysRevLett.79.325}. We find that 
in BB84 protocol, using dissension vector, the respective parties can detect the presence of 
an eavesdropper. In the case of Grover search algorithm, we find that to achieve success in 
Grover search algorithm, a substantial amount of dissension should develop during the 
processes. The dissension vectors can trace the correlations, and are maximum where 
one expects maximum correlations.

The organization of the paper is as follows. In Section 2, we discuss classical 
mutual information and its
extension to quantum regime. We discuss correlations and quantumness in Section 3.
In Section 4, we  extend the notion of discord along two different tracks and 
give expressions for dissension vectors for $n$-qubit case.
In Section 5, we analyze these measures with examples for three- and four-qubit systems. 
We discuss some features of these measures in Sections 6.  In section 7, we address the possible 
physical applications of these measures.
Finally we conclude in Section 8.
\section{Mutual information and its generalization to Quantum regime}
Let us consider two random variables $X$ and $Y$. The common information that they possess 
is characterized by mutual information
\begin{equation}
 I(X:Y)=H(X)+H(Y)-H(X,Y),
\end{equation}
 where $H(X)$ is Shannon entropy of $X$ and $H(X,Y)$ is the joint entropy. 
There are many  uses of mutual information. Our interest is in its ability
to capture correlations between two probability distributions. Using chain 
rule, one can express mutual information also as,
\begin{eqnarray}
 I(X:Y)&=& H(X)-H(X|Y),\nonumber\\
 &=&H(Y)-H(Y|X),\nonumber\\
 &=&H(X,Y)-(H(X|Y)+H(Y|X)),
\label{alterI}
\end{eqnarray}
where $H(X|Y)=H(X,Y)-H(X)$ is the conditional entropy. In Eq.(\ref{alterI}), the last expression 
of mutual information is symmetric in $X$ and $Y$ unlike the former two. Note that the quantity 
$H(X|Y)+H(Y|X)$ is metric in its own right and called `variation of information'.

In quantum regime, mutual information is written in terms of 
von Neumann entropy of density matrices. Intuitively this 
quantity solely should characterize the correlations between two subsystems 
of a bipartite system. But in reality it does not. It is 
sometimes suggested that the mutual information quantifies the total 
correlations of a bipartite system \cite{PhysRevA.72.032317}. However, in general what 
it characterizes about the state is somewhat elusive \cite{2009PhLA..373.1818W, PhysRevA.76.032327}. Also the 
generalization of this quantity to quantum regime leads  to many 
new features and complexities. One way of generalization is 
that of replacing the probability distributions with density matrices and 
another is using relative entropy, i.e., for a bipartite state $\rho_{xy}$,
\begin{eqnarray}
 I^q(X:Y)&=& S(X)+S(Y)-S(X,Y),\nonumber\\
 &=&S(\rho_{XY}\parallel\rho_X\otimes\rho_Y),
\label{qi-2}
\end{eqnarray}
where $S(X)=-\tr(\rho_X \log_2 \rho_X)$ represents von Neumann entropy and 
$S(\rho\parallel\sigma)=\tr\rho(\log_2\rho-\log_2\sigma)$ is relative entropy. 
\subsection{Quantum conditional entropy and mutual information}
possible generalizations of Eq. (\ref{alterI}) for the bipartite quantum state $\rho_{XY}$ are,
\begin{eqnarray}
 I_Y(X:Y)&=& S(X)-S(X|Y),\nonumber\\
 I_X(X:Y)&=&S(Y)-S(Y|X),\nonumber\\
 I_a(X:Y)&=&S(X,Y)-(S(X|Y)+S(Y|X)),
\label{alterQI}
\end{eqnarray}
where $S(X|Y)$ is the quantum conditional entropy. If we directly 
extend the classical conditional entropy expression to quantum domain, then 
$S(X|Y)=S(X,Y)-S(Y)$, which is negative for 
pure entangled states. This negativity of conditional entropy was 
explained in the references 
\cite{PhysRevLett.79.5194, PhysRevA.60.893,2005Natur.436..673H, nature.delN, Horodecki2007}. However, there is an alternate
view which says that to know a state we have to make
a measurement \cite{discordHend2001}. This is then the meaning of ``conditional".
So, conditional entropy can also be expressed as,
\begin{equation}
 S(X|Y)=\displaystyle\sum_i p_iS(\rho_{X|\Pi_i^Y}),
 \label{Cond_entr}
\end{equation}
where $\rho_{X|\Pi_i^Y}=\frac{1}{p_i}\tr_Y(\mathbb{I}_2\otimes\Pi_i^Y)\rho_{XY}(\mathbb{I}_2\otimes\Pi_i^Y)$ 
with $p_i=\tr(\mathbb{I}_2\otimes\Pi_i^Y)\rho_{XY}(\mathbb{I}_2\otimes\Pi_i^Y)$. $\mathbb{I}_p$ 
is the identity matrix of order $p$ and $\{\Pi_i^Y;\hspace{0.1cm}i=1,2\}$ are, in general,
the rank one positive operator valued measure (POVM) on part $Y$. The definition in Eq.(\ref{Cond_entr}) 
is always positive.

\subsection{Multiparticle mutual information}

Our goal in this paper is to examine multiparticle systems. So we need 
a generalization of the bipartite mutual information to a multipartite 
situation. We will use the usual generalization based on Venn diagram 
approach. In this approach, the mutual information for 
three variables $X$, $Y$ and $Z$ is defined as
\begin{equation}
 I_0(X:Y:Z)=I(X:Y)-I(X:Y|Z),
\label{cond-Im}
\end{equation}
where $I(X:Y|Z)=H(X|Z)+H(Y|Z)-H(X,Y|Z)$ is conditional mutual information \cite{cover1991}. This can be immediately
generalized to $n$-variate mutual information. Using chain rules, this generalization
will lead to the multivariate mutual information as,
\begin{eqnarray}
 I_0(X_1:..:X_n)=\displaystyle\sum_{p=1}^n(-1)^{p-1}\displaystyle\sum_{\{l_p\}}^nH(X_{l_1},X_{l_2},...,X_{l_p}),
\end{eqnarray}
where $\{l_p\}$ in the sum denotes that if $p=k$, then indices $l_1,l_2,...,l_k$ ($k\leq n$) will survive with each $l_i$ varies from $1$ to $n$ and $l_i\neq l_j$. 
In literature, this quantity is also known as the `interaction information'. 
By analogy, one can write the multiparticle mutual information of the 
state $\rho_{x_1x_2..x_n}$ as,
\begin{eqnarray}
 I_0^q(x_1:x_2:...:x_n)=\displaystyle\sum_{p=1}^n(-1)^{p-1}\displaystyle\sum_{\{l_p\}}^nS(x_{l_1},x_{l_2},..,x_{l_p}),
\label{muI}
\end{eqnarray} 
where $x_i$ here stands for $x_i^{th}$ subsystem. 
This generalization has not been explored much. In this paper, we will use this generalization 
and define a vector type correlation measure to characterize and quantify multiparticle correlations.

However, there exist at least two more mutual information like quantities in literature. First one is 
known as `total correlation'. The total correlation for three variables is 
\begin{equation}
I_t(X:Y:Z)=I(X:Y)+I(XY:Z),
\end{equation}
where $I(XY:Z)=I(X:Z)+I(Y:Z|X)$. This quantity can also be generalized for multi-variables i.e.,
\begin{eqnarray}
I_t(X_1:...:X_n)=\displaystyle\sum_{i=1}^nH(X_i)-H(X_1,...,X_n).\label{total_I}
\end{eqnarray}
It can be generalized to quantum regime for the state $\rho_{x_1x_2...x_n}$
\begin{eqnarray}
 I_t^q(x_1:x_2:...:x_n)
&=&\displaystyle\sum_{p=1}^nS(x_i)-S(x_1,x_2,...,x_n)\nonumber\\
&=&S(\rho_{x_1,x_2,...,x_n}\parallel\displaystyle\otimes_{i=1}^n\rho_{x_i}).
\label{GI-unu}
\end{eqnarray}
The second line of the Eq.(\ref{GI-unu}) shows that it is a distance between the state 
and tensor products of its marginals. 
This generalization has been used in literature \cite{PhysRevA.72.032317} to capture total correlations in a 
multiparticle quantum state. 
Note that the above generalization is always positive \cite{MI-Herbut2004}. 

\begin{figure}[!htb]
\begin{minipage}{0.45\textwidth}
\centering
\[
\begin{array}{ccc}
\includegraphics[height=4.8cm,width=5cm]{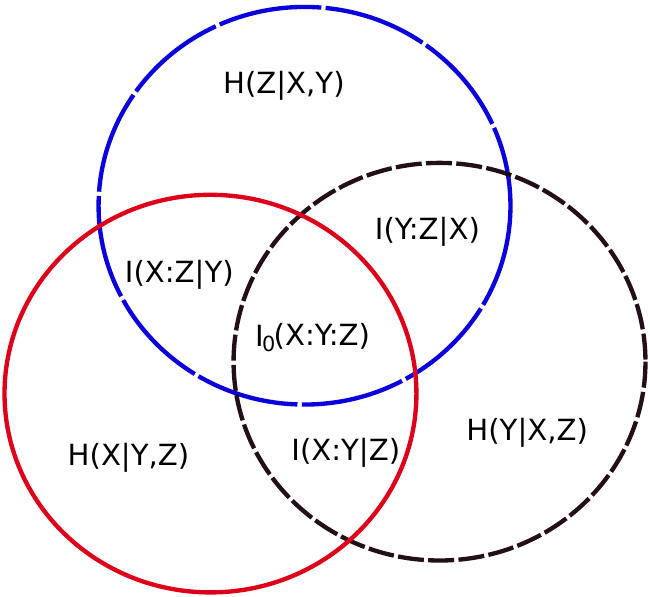}
\end{array}
\]
\caption{(Color online) \textcolor{blue}{Venn diagram}: The information theoretic quantities for three random variables, $X$, $Y$, and $Z$. The total 
correlation, $I_t(X:Y:Z)=I_b(X:Y:Z)+I_0(X:Y:Z)$, and the binding information, $I_b(X:Y:Z)=I(X:Y|Z)+ I(X:Z|Y)+I(Y:Z|X)+I_0(X:Y:Z)$, 
where $I_0(X:Y:Z)$ is the interaction information.}
\label{3mutualv}
\end{minipage}
\hspace{0.5cm}
\begin{minipage}{0.45\textwidth}
\centering
\[
\begin{array}{ccc}
\includegraphics[height=4.8cm,width=7.5cm]{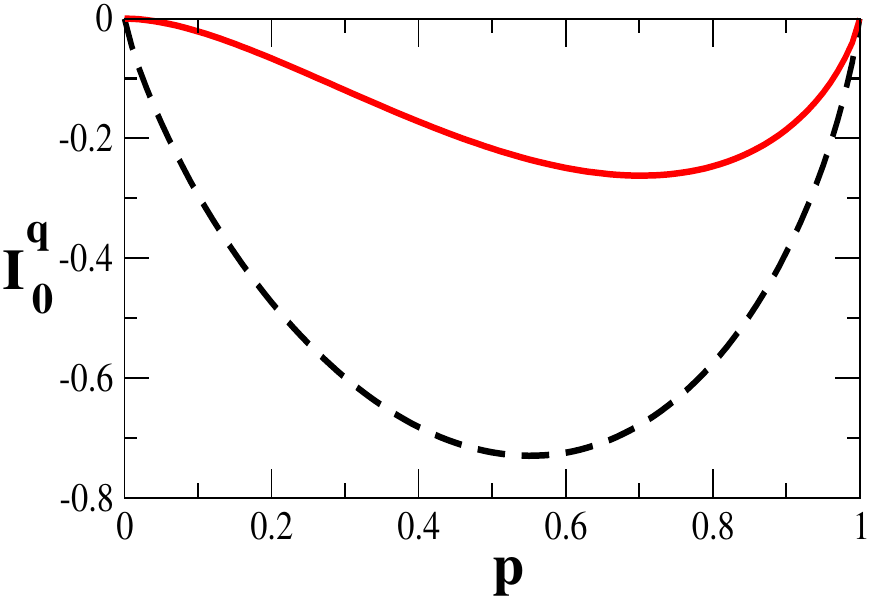}
\end{array}
\]
\caption{(Color online) Figure shows the plot of interaction information $I_0^q$ with the mixing parameter $p$ for the 
three qubit mixed states $\rho_{GW}=(1-p)\ketbra{W_3}{W_3}+p\ketbra{G_3}{G_3}$ (black dashed line) and 
$\rho_{Wer}=\frac{(1-p)}{8}\mathbb{I}_8+p\ketbra{G_3}{G_3}$ (solid red line). It shows interaction information 
can be negative for three qubit states.}
\label{3qI0q}
\end{minipage}
\end{figure}

Another quantity is the `dual total correlation', or `binding information' or, sometime known as `secrecy monotone' 
\cite{PhysRevA.66.042309}. For three random variables it is expressed as 
\begin{equation}
I_b(X:Y:Z)=I(X:YZ)+I(Y:Z|X),
\end{equation}
where $I(X:YZ)=I(X:Y)+I(X:Z|Y)$. The above quantity can be generalized for multi-variables i.e.,
\begin{eqnarray}
I_b(X_1:...:X_n)=\sum_{i=1}^nH(X_1,...,X_{i-1},X_{i+1},...,X_n)-(n-1)H(X_1,...,X_n).
\end{eqnarray}
The quantity, $I_b(x_1:x_2:...:x_n)$ can easily be extended for the multiparticle quantum state, $\rho_{x_1x_2...x_n}$ 
\begin{eqnarray}
I_b^q(x_1:x_2:...:x_n)=\sum_{i=1}^nS(x_1,...,x_{i-1},x_{i+1},...,x_n)-(n-1)S(x_1,...,x_n).
\label{secrecy_mono}
\end{eqnarray}
Note that the above quantity is also always positive \cite{2015arXiv150407176K} and for pure states 
$I_t^q(x_1:x_2:...:x_n)=I^q_b(x_1:x_2:...:x_n)$. This quantity has been used in literature for capturing 
correlations in a quantum state \cite{2015arXiv150407176K} and to detect the shared secret correlations between the parties 
\cite{PhysRevA.66.042309}. The total correlation, $I_t^q(x_1:x_2:...:x_n)$ and the 
binding information, $I^q_b(x_1:x_2:...:x_n)$ are monotones under CPTP (complete positive trace preserving) map
\cite{PhysRevA.66.042309,2015arXiv150407176K}. Moreover, for two particle quantum systems, 
the Eqs.(\ref{muI}, \ref{GI-unu} and \ref{secrecy_mono}) reduce to $I^q(x_1:x_2)$. 
The Fig.(\ref{3mutualv}) depicts the relations between the possible generalizations of multivariate 
mutual information. These relations may not hold for quantum case. From the diagram, it is clear that
only $I_{0}$ characterizes genuine multiparticle correlations. Other two generalizations,
$I_t$ and $I_{b}$ also contain bipartite correlations. 

An important feature of interaction information is: Negative of tripartite quantum interaction information ($I_{0}^q$) 
is useful in determining the achievable rate for a particular secret sharing task, the information scrambling 
\cite{JHEP145,2017arXiv170302903S}. If interaction information is negative then one can conclude that 
achievable rate will be nonzero for such a secret sharing task. If we carefully inspect the Table.\ref{CompIoIsI}, it shows that $I_{0}^q$ may not capture total or genuine correlations in the quantum state, but it may be useful in characterizing 
certain highly entangled states. The Fig.(\ref{3qI0q}) indicates that $I_{0}^q$ can be negative for three qubit mixed states. 
We are using generalization of $I_{0}$ to quantum domain.

\begin{table}
\begin{center}
\begin{tabular}{|c|c|c|c|}
\hline 
State & $I_0^q$ & $I_b^q$ & $I_t^q$\\
\hline $\ket{G_4}$ & 2 & 4 & 4\\
\hline $\ket{C}$ & -2 & 4 & 4\\
\hline $\ket{HS}$ & -2.755 & 4 & 4\\
\hline $\ket{W_4^2}$ & 0.49 & 4 & 4\\
\hline $\ket{G_2}^{\otimes 2}$ & 0 & 4 & 4\\
\hline
\end{tabular}
\captionof{table}{$I_0^q$ vs $I_t^q$ and $I_b^q$ table. 
For four qubit 
case, $I_0^q$ may be used to characterize different highly entangled states which may not be 
possible using the other two. 
The states are $\ket{C}=\frac{1}{2}(\ket{0000}+\ket{0011}+\ket{1100}-\ket{1111})$, 
$\ket{HS}=\frac{1}{\sqrt{6}}(\ket{0011}+\ket{1100}+
\omega\{\ket{1010}+\ket{0101}\}+\omega^2\{\ket{1001}+\ket{0110}\})$, 
$\ket{W_n^r}=1/\sqrt{\binom{n}{r}}(\sum_{\mathcal{P}}\mathcal{P}\ket{0}^{\otimes n-r}\ket{1}^{\otimes r})$, 
$\ket{G_n}=\frac{1}{\sqrt{2}}(\ket{0}^{\otimes n}+\ket{1}^{\otimes n})$.  
Here $\mathcal{P}$ denotes all possible combinations, $\omega=e^{\frac{2\pi i}{3}}$.
}
\label{CompIoIsI}
\end{center}
\end{table}

\subsection{Can Mutual Information be negative?}

One feature of the multivariate mutual information, as given by Venn diagram
approach, is that it can be negative. Sometimes, it is considered a negative
aspect of this approach. However, as we will see, the negative value characterizes
a very special type of correlations \cite{cover1991}. For this we consider mutual information 
of three variables $X$, $Y$ and $Z$, as given in Eq.(\ref{cond-Im}). In this
definition, both
$I(X:Y)$ and $I(X:Y|Z)$ are non-negative, but $I_0(X:Y:Z)$ can be negative, 
when $I(X:Y)<I(X:Y|Z)$. This situation will occur when knowing $Z$ enhances 
the correlation between $X$ and $Y$. Let us take a well known example of 
`modulo 2 addition ($\oplus$) of two binary random variables (XOR-gate)'. Suppose, 
$X\oplus Y=Z$. If $X$ and $Y$ are independent then $I(X:Y)=0$. However, 
once we know the value of $Z$, knowing the value of $X$ uniquely 
determines the value of $Y$. Hence the knowledge of $Z$ enhances the 
correlation between $X$ and $Y$, i.e., $I(X:Y|Z)$ is non-zero. 
This implies when $I_0(X:Y:Z)$ is negative, it captures certain aspect 
of the correlations among the variables $X$, $Y$ and $Z$.

The generalization of Eq. (\ref{cond-Im}) in the quantum regime, for the state $\rho_{XYZ}$ is 
\begin{equation}
 I^q_0(X:Y:Z)=I^q(X:Y)-I^q(X:Y|Z),
 \label{cond-Imq}
\end{equation}
where $I^q(X:Y|Z)=S(X|Z)+S(Y|Z)-S(XY|Z)$ 
is conditional mutual information. Let us consider the case of a three-qubit GHZ state 
$\ket{G_3}=\frac{1}{\sqrt{2}}(\ket{000}+\ket{111})$. If we trace out any 
one qubit from the state then the reduced density matrix is a mixture of 
product states, i.e., $\rho_{r}=\frac{1}{2}(\ket{00}\bra{00}+\ket{11}\bra{11})$. 
For this state, the mutual information is $I^q(X:Y)=1$. Now, the conditional mutual information, 
$I^q(X:Y|Z)$ for $\ket{G_3}$ will depend on the measurement basis. 
We know $S(XY|Z)=0$ in any measurement basis but it is not the case for other two terms 
$S(X|Z)$ and $S(Y|Z)$. The measurement on qubit $Z$ in computational basis ($\{\ket{0},\ket{1}\}$) will 
give $S(X|Z)=S(Y|Z)=0$, i.e., $I^q(X:Y|Z)=0$. So the total mutual information is 
$I_0^q(X:Y:Z)=1$, i.e., positive. It is not surprising because the 
state of remaining two qubits, after measurement on one qubit, does not
have enhanced entanglement. But for the measurement on qubit $Z$ in 
Hadamard basis ($\{\ket{+},\ket{-}\}$), the mutual information, $I^q(X:Y|Z)=2$, 
which means, total mutual information is $I^q_0(X:Y:Z)=-1$, i.e., negative.
This is expected, since now the state of two qubits $XY$ is a Bell state; so measurement
on $Z$ qubit has enhanced the entanglement in $XY$ subsystem.
The essence of this discussion is that in both classical and 
quantum regime multivariate mutual information can be negative, characterizing a
special type of correlations.  

Let us re-express the Eqs.(\ref{cond-Im} \& \ref{cond-Imq}) such that we find the following compact expression
\begin{equation}
 I_0(X:Y:Z)=I(X:Y)+I(X:Z)-I(X:YZ).
\end{equation}
The above expression is positive if $I(X:Y)+I(X:Z)\geq I(X:YZ)$, i.e., when mutual information is monogamous. 
Otherwise it is polygamous. Hence, a negative $I_0(X:Y:Z)$ means that the correlations between $X$ and the 
joint system $YZ$ is more than the sum of the individual ones. It is well understood that the entanglement among other correlations is always monogamous in nature while classical correlations is not. However, presence of 
strong entanglement as well as classical correlations between $X$ and joint system $YZ$ makes the situation complicated.
\section{Correlations and Quantumness}

Whether a quantum state (of more than one particle) has correlations 
or not, it is often far from obvious. This is because the meaning of the 
word `correlation', as often used in literature, is quite fluid. 
We know the meaning of correlation in classical world 
but in the case of a quantum state there are classicality 
and quantumness. This makes the nature of correlations very complex. 
If we take the intuitive meaning of correlations \cite{RevModPhys.81.865}, then
quantum correlations are nonlocal in nature, and can be taken
as due to entanglement of the state only.  They exist due to
the nonlocal quantumness of a state. A state can also have classical
correlations \cite{discordHend2001} and local quantumness \cite{2015arXiv150200857A}. When we speak of quantumness
of a state, it can be local or nonlocal in character. Information
theoretic measures like quantum discord, and its generalization
like dissension, characterize and quantify both types of quantumness.
Next we emphasize the need of a vector measure to characterize
the quantumness of a state. We then expand on local and nonlocal
quantumness. 

\subsection{Quantum Discord: Is one number sufficient?}

In the reference \cite{PhysRevLett.88.017901, discordHend2001}, authors have given a way of quantifying 
quantum correlations present in bipartite two-qubit states through quantum 
discord. To do so they used different generalizations of the mutual information 
to quantum regime. Let us consider the bipartite state $\rho_{xy}$. 
Then using Eqs. (\ref{qi-2}) and (\ref{alterQI}), 
the discord is defined in the following way,
\begin{equation}
 \delta_j (\rho_{xy})=\inf_{\Pi^j}\{I^q_0(x:y)-I_j(i:j)\},
\label{vec-xy}
\end{equation}
where $I_j(i:j)=S(i)-S(i|j)$ with $\{i,j;\hspace{0.1cm} i\neq j\}=x,y$. Here, the measurement bases, $\Pi^j$ 
are $\{\cos\theta \ket{0}+e^{i\phi}\sin\theta \ket{1},-\sin\theta\ket{0}+e^{i\phi}\cos\theta\ket{1}\}$ 
with $\theta \in [0,\pi]$ and $\phi \in [0,2\pi]$. (Note that throughout the manuscript, we have considered 
this basis as our single particle measurement basis.)
Obviously the above definition is not symmetric in the parties. When $j=y$, it is usual 
discord and for $j=x$ it is $\delta_x(\rho_{xy})$. Sometimes one of the discords is 
zero, even when other is nonzero. If we take that the quantum discord captures 
`quantumness' present in a state, it is quite clear that we need both the discords to know 
the exact quantumness of the state.

One can define `another discord' as, 
\begin{equation}
 \delta_a(\rho_{xy})=\inf_{\{\Pi^x,\Pi^y\}}\{I^q_0(x:y)-I_a(x:y)\}.
\label{vec-x+y}
\end{equation}
This definition is symmetric in the parties.\\ 
\noindent \textbf{Lemma.1} \textit{The quantity, $\delta_a(\rho_{xy})$, is nothing but the 
sum of the two discords $\delta_x(\rho_{xy})$ and $\delta_y(\rho_{xy})$}.\\ 
{\emph Proof:} Using Eq.(\ref{alterQI}) we have 
\begin{eqnarray}
 \delta_a(\rho_{xy})&=&\inf_{\{\Pi^x,\Pi^y\}}\{I^q_0(x:y)-I_a(x:y)\}\nonumber\\
 &=&\inf_{\{\Pi^x,\Pi^y\}}\{S(x)+S(x|y)-S(xy)+S(y)+S(y|x)-S(xy)\}\nonumber\\
 &=&\inf_{\Pi^y}\{I^q_0(x:y)-I_y(x:y)\}+\inf_{\Pi^x}\{I^q_0(x:y)-I_x(x:y)\}\nonumber\\
 &=&\delta_x (\rho_{xy})+\delta_y (\rho_{xy}).\nonumber
\end{eqnarray}
Hence proved. $\square$

Let us compute the above quantities for the following examples. For this 
purpose we introduce a vector-type quantity $\{\delta_x,\delta_y\}$ 
instead of using $\delta_x$ and $\delta_y$ separately. It is a quantumness 
vector -- discord vector. For example, consider the following two-qubit states. Both for product states, 
$\rho_p=\ket{\phi}\otimes\ket{\chi}$ and classically 
correlated states, $\rho_c=p\ketbra{\phi\chi}{\phi\chi}+p\ketbra{\phi^{\perp}\chi^{\perp}}{\phi^{\perp}\chi^{\perp}}$, 
the discords are  $\{\delta_x,\delta_y\}=\{0,0\}$, $\delta_a=0$, where $\bra{\phi}\phi^{\perp}\rangle=0$. 
The classical-quantum state, $\rho_{cq}  =  \frac{1}{2}(\ketbra{++}{++}+\ketbra{-0}{-0})$ and the quantum-classical states,  
$\rho_{qc}  =  \frac{1}{2}(\ketbra{++}{++}+\ketbra{0-}{0-})$ have the discords $\{\delta_x,\delta_y\}=~\{0,0.2\}$ 
and $\{\delta_x,\delta_y\}=~\{0.2,0\}$ respectively but the $\delta_a=0.2$ for both the states, 
where $\ket{\pm}=\frac{1}{\sqrt{2}}(\ket{0}+\ket{1})$. Whereas a quantum-quantum 
separable state, $\rho_{qq}=\frac{1}{2}(\ketbra{00}{00}+\ketbra{++}{++})$ have $\{\delta_x,\delta_x\}=\{0.15,0.15\}$, $\delta_a=0.3$. 
As expected, the maximally entangled state, $\ket{G_2}=\frac{1}{\sqrt{2}}(\ket{00}+\ket{11})$ will have discords 
$\{\delta_x,\delta_x\}=\{1,1\}$, $\delta_a=2$.
Hence, the vector type quantification of correlation reveals more information about the 
correlation of a state than $\delta_x$ or $\delta_y$ alone. 
\subsection{Local and nonlocal quantumness}
As has been argued in \cite{2015arXiv150200857A}, the quantities like quantum discord
not only characterize nonlocal quantumness (i.e., entanglement),
but also local quantumness. Same will continue to hold for the
generalizations of the discord that we will discuss. Let us recall
these aspects of quantumness.

Let us consider the example states from the previous subsection.
We found that the both $X$-discord 
and $Y$-discord are zero for $\rho_p$ and $\rho_c$. But $X$-discord is zero and 
$Y$-discord is non-zero for $\rho_{cq}$ and both discords are non-zero for $\rho_{qq}$. 
Now consider the state, $\rho_p$. In Hadamard basis, there is no superposition 
in it but in computational basis there is. So one can mask the local superposition 
of the state $\rho_p$ and hence discord is zero. State $\rho_c$ is the mixture 
of orthogonal states and there is no local quantumness in it. But the states $\rho_{cq}$ 
and $\rho_{qq}$ are the mixture of non-orthogonal states. State $\rho_{cq}$ has local 
superposition in one part and that's why one of the discords is non-zero for this case 
but for $\rho_{qq}$ both parts have local superposition. Formally one can define local 
quantumness as following.\\
{\bf Definition:} {\em We say that a mixture of non-orthogonal separable states has local quantumness 
(i.e., local superposition), if it cannot be masked by writing down the state in another 
decomposition.} 

\begin{figure}[!htb]
\begin{minipage}{0.45\textwidth}
\centering
\[
\begin{array}{ccc}
\includegraphics[height=4.8cm,width=7.5cm]{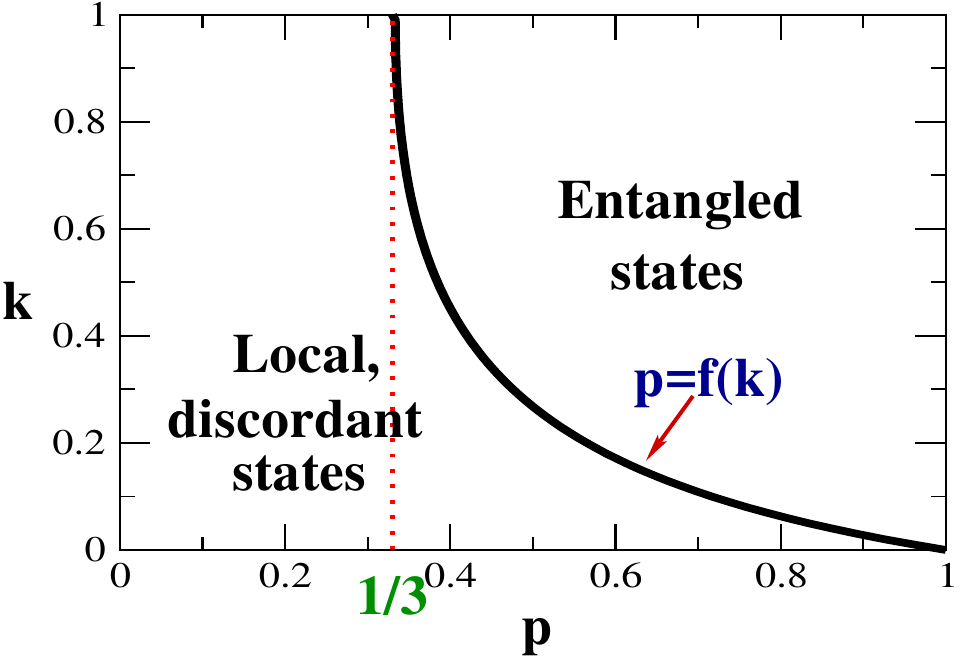}
\end{array}
\]
\caption{(Color online) \textcolor{blue}{Phase diagram}: The figure depicts phase diagram of two qubit states. 
The black solid curve is the $p=f(k)$ curve, where $f(k)$ = $(1+k^2/(1+4k+k^2)$. The states on the right hand side of black 
curve are entangled ones and on the left side almost all states have local quantumness 
(separable states with non-zero discord) except the $k=0$ line.}
\label{phasedde}
\end{minipage}
\hspace{0.5cm}
\begin{minipage}{0.45\textwidth}
\centering
\[
\begin{array}{ccc}
\includegraphics[height=5cm,width=7cm]{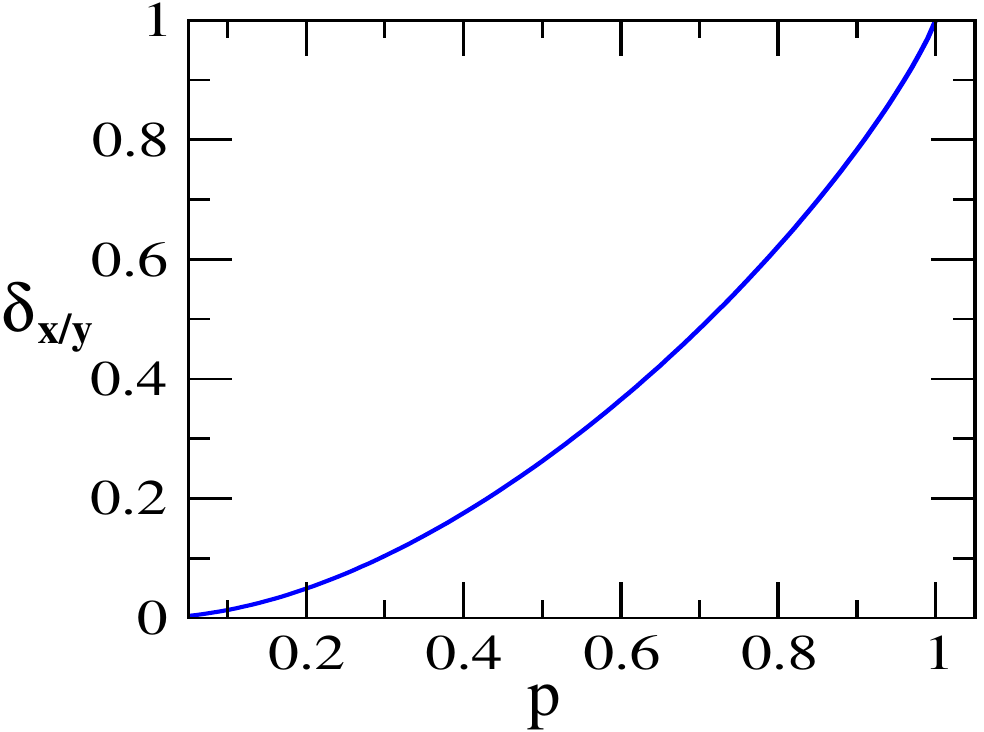}
\end{array}
\]
\caption{(Color online) The figure shows how the $\delta_x(\delta_y)$ 
behaves as a function of mixing parameter $p$ for the Werner state, $\rho_{Wer}$. Note that the state 
has non-zero discord for $p\leq \frac{1}{3}$ when it has no entanglement. This shows this discord is due to 
local quantumness.}
\label{comwer}
\end{minipage}
\end{figure}

To illustrate the various features of a state, 
let us consider the generalized Werner state \cite{2015arXiv150200857A},
\begin{equation}
 \rho_{Wg}=\frac{(1-p)}{4}\mathbb{I}_4+p\rho_{k},
\label{gen-wer}
\end{equation}
where $\rho_{k}=\ket{\psi}_{k}\bra{\psi}$ with 
$\ket{\psi}_{k}=\frac{1}{\sqrt{1+k^2}}(\ket{00}+k\ket{11})$. Here $p$ is 
classical mixing parameter, whereas $k$ is the nonlocal parameter due to its
role in nonlocal superposition. 
The state (see Eq.(\ref{gen-wer})) 
is separable if $p\leq\frac{1+k^2}{1+4k+k^2}$ but discord is non-zero. 
Nonetheless, one can show that the above state (see Eq.(\ref{gen-wer})) has both local and 
nonlocal quantumness (see Fig.(\ref{phasedde})). For $p\leq\frac{1+k^2}{1+4k+k^2}$, one rewrite the state as valid 
mixture of non-orthogonal states, and the state has only local quantumness. 

For example, if we consider $k=1$, the state coincides with the Werner state, 
$\rho_{Wer}=\frac{(1-p)}{4}\mathbb{I}_4+p\rho_{G_2}$.  
The Fig.(\ref{comwer}) depicts the behavior of both the discords with mixing parameter $p$ for 
the two-qubit Werner state. The Werner state is separable for $p\leq\frac{1}{3}$. 
This is because one can always rewrite Werner state in such a 
way that the state is a valid mixture of non-orthogonal states whenever  $p < {1 \over 3}$ \cite{PhysRevA.66.022104}. 
Rewriting the Werner state in that form, we have
\begin{eqnarray}
 \rho_{\rm Wer}= (1 - 3 p)\,\frac{I}{4} 
  +\frac{p}{2}\, (|++\rangle\langle ++| + |--\rangle\langle --|+ 
|00\rangle\langle 00| +|11\rangle\langle 11|   
+|\tilde{+}\tilde{-}\rangle\langle \tilde{+}\tilde{-} |
 +  |\tilde{-}\tilde{+}|\rangle\langle |\tilde{-}\tilde{+}| ), 
\end{eqnarray}
 where $ | \tilde{\pm} \rangle\ = {1 \over \sqrt{2}}  (|0\rangle \pm i |1\rangle) $. 
 This decomposition is valid only for $ p \leq \frac{1}{3}$. This is 
 precisely the region of $p$, where Werner  state is not entangled.
 Since $ \langle +| 0 \rangle \neq 0$ and  $ \langle +| \tilde{+} \rangle \neq 0$, this 
 state is a mixture of separable non-orthogonal states; so it is expected
 to have non-zero discord due to local quantumness.
\section{Dissension vectors}
From the discussion of the last section, it is clear that a 
vector-type correlation measure is 
better in describing the quantum properties of a state. Using 
multivariate mutual information, we will now generalize the 
quantum discord to $n$-qubit system, calling it dissension. 
We will introduce two types of quantumness vectors -- called dissension vectors.

Let us consider a state $\rho_{x_1x_2...x_n}$ in Hilbert space 
$H_2\otimes H_2\otimes...\otimes H_2$ where $x_i$ qubit is with 
$i^{th}$ party. The mutual information for this state is (see Eq. \ref{muI})
\begin{eqnarray}
 I^q_0(x_1:x_2:...:x_n)=\displaystyle\sum_{p=1}^n(-1)^{p-1}\displaystyle\sum_{\{l_p\}}^nS(x_{l_1},x_{l_2},..,x_{l_p}).
\label{muIxc}
\end{eqnarray}
Using chain rule, we can now introduce
conditional entropies. These conditional entropies are to be understood in
terms of measurements. In this way, one can introduce one party, 
two party, ......, ($n-1$)-party measurements in the above expression of mutual 
information (see Eq.(\ref{muIxc})) and each leads one to a expression for new mutual information. When more 
than one party is involved, joint measurement is to  be implemented. Following 
reference \cite{Chakrabarty2011}, one can have mutual information with all possible  
conditionals which we called Track-II type definition, but following \cite{PhysRevLett.88.017901, discordHend2001} 
one can have mutual information with smaller number of conditionals which we call 
Track-I (or discord track) definition of mutual information.

\begin{quote}
{\em Notation}.- Before defining the dissension vectors, we want to clarify what we mean by the notation 
$\vec{\delta}_m^t$. The index $m$ defines the number of particles on which the joint measurement has been 
performed and $t$ indicates the track.
\end{quote}

\subsection{Track-I}
Let us consider the most general situation where we have state $\rho_{x_1x_2...x_n}$ 
with $n$ number of qubits. On the basis of $m$-party joint measurement (One can employ local measurements simultaneously.), we will have 
$(n-1)$ expressions for mutual information 
$\{I_m^1(x_1:x_2:...:x_n);m=1,2,...,(n-1)\}$,
\begin{eqnarray}
I_m^1(x_1:x_2:...:x_n)&=&\displaystyle\sum_{k=1}^{m-1}(-1)^{k-1}\displaystyle\sum_{\{l_k\}}^nS(x_{l_1},x_{l_2},..,x_{l_k})
 +(-1)^{m-1}\displaystyle\sum_{\{k_{m-1}\};k_1=2}S(x_1,x_{k_1},...,x_{k_{m-1}})
 \nonumber\\&+&\displaystyle\sum_{p=m+1}^n(-1)^{p-1}\displaystyle\sum_{\{l_p\}}^nS(x_{l_1},..,x_{l_{p-m}}
 |x_{l_{p-m+1}},..,x_{l_p})
\label{genmu}
\end{eqnarray}
where, $S(x_{l_1},..,x_{l_{p-m}}|x_{l_{p-m+1}},..,x_{l_p})$ denotes conditional 
entropy where joint measurement are to be  done on parties $x_{l_{p-m+1}},..,x_{l_p}$. 
We can define dissension function, $D_m^1(\rho_{x_1x_2...x_n})=(-1)^n(I^q_0-I_m^1)$, and then the dissension, 
\begin{eqnarray}
 \delta_m^1=\inf_{\Pi_m}[(-1)^n(I^q_0-I_m^1)],
\end{eqnarray}
where minimization is done over $m$-party measurement. The expressions of mutual information in Eq.(\ref{genmu}) are not symmetric 
under interchange of parties. For example, if we take $m=1$, $I_1^1$ can have $n$ 
number of different expressions which are very different from one another. In 
Eq. (\ref{genmu}), if we put $m=1$, we can have one type of $I_1^1$; let us 
name it $I_{x_n}^1$. Now exchanging $x_n$ with $x_1,x_2,.....,x_{n-1}$ 
respectively one can have others. So we have $n$ number of $\delta_1^1$. We 
label them as $\delta_{x_p}^1=(-1)^n(I^q_0-I_{x_p});p=1,2,...,n$. This leads us 
to define dissension vector
\begin{eqnarray}
 \vec{\delta}_1^1=\{\delta_{x_p}^1;p=1,2,...,n\}.
\end{eqnarray}
In this way with some particular choice of entries one can have $n-1$ vectors i.e., 
$\vec{\delta}_i^1;i=1,2,...,(n-1)$.
\subsection{Track II}
Next we extend  the definitions of mutual information in this track to all possible 
$m$ party conditionals and we have the expression for mutual information i.e., 
$I_m^2;m=1,2,....,(n-1)$,
 \begin{eqnarray}
 I_m^2&=&\displaystyle\sum_{k=1}^{m-1}(-1)^{k-1}\displaystyle\sum_{\{l_k\}}^nS(x_{l_1},x_{l_2},..,x_{l_k})+(-1)^{m-1}
 \left[\displaystyle\sum_{\{k_{m-1}\};k_1=2}^{n-1}\{S(x_1,x_{k_1},...,x_{k_{m-1}},x_{k_{m-1}+1})
  \nonumber\right.\\&-&\left. S(x_{k_{m-1}+1}|x_1,x_{k_1},...,x_{k_{m-1}})\}+\cdot\cdot\cdot\cdot\cdot\cdot\cdot
 + S(x_1,x_2,x_{k_{n-m+2}},x_{k_{n-m+3}},...,x_{n-1},x_{n})
 \nonumber\right.\\
 &-&\left.S(x_2|x_1,x_{k_{n-m+2}},...,x_{n-1},x_{n})\right]
 +\displaystyle\sum_{p=m+1}^n(-1)^{p-1}\displaystyle\sum_{\{l_p\}}^nS(x_{l_1},..,x_{l_{p-m}}
 |x_{l_{p-m+1}},..,x_{l_p}).
\label{2-gm}
\end{eqnarray}

The dissension function in this track is defined as $D_m^2=(-1)^n(I^q_0-I_m^2)$. Therefore, the dissensions are 
\begin{eqnarray}
 \delta_m^2=\inf_{\Pi_m}[(-1)^n(I^q_0-I_m^2)].
\end{eqnarray}
If we interchange parties, the mutual information in the Eq.(\ref{2-gm}) will not 
remain same except for $m=(n-1)$. For example if we consider $m=1$ in the Eq.(\ref{2-gm}), 
we will get one $I_1^2$; let us call it as $I_{x_n}^2$. Now interchanging
 $x_n$ with $x_1,x_2,....,x_{n-1}$ respectively we will get others. In this way, we 
will have $n$ numbers of $\delta_1^2$. If we label them as 
$\delta_{x_p}^2=(-1)^n(I^q_0-I_{x_p}^2);p=1,2,...,n$, we have dissension vector
\begin{eqnarray}
 \vec{\delta}_1^2=\{\delta_{x_p}^2;p=1,2,...,n\}.
\end{eqnarray}
With some particular choice of entries one can have $n-2$ vectors i.e., 
$\vec{\delta}_i^2;i=1,2,...,(n-2)$ and one symmetric quantity $\delta_{n-1}^2$. We call 
these quantities dissension vectors in Track-II.

\section{Simple illustrations}

In this section, we will present our numerical results for a set of 
three-qubit and four-qubit states. It will illustrate the 
usefulness of the dissension vectors. We will consider track-I 
and track-II dissension vectors, as defined in the last section. We will 
see that both tracks are most of the time useful.

\subsection{Three-qubit states}
For the three qubit states, the dissensions have been extensively calculated in the work \cite{Chakrabarty2011}. For the sake of 
completeness, we have analyzed the dissension vectors for three qubit states and discuss some of the results. For three 
qubits, the dissension vectors are ($\vec{\delta}_1^1$, $\vec{\delta}_2^1$) in track-I and in track-II, $\vec{\delta}_1^2$ along with the 
symmetric discord $\delta_2^2$. Here in this work, we consider two qubit joint measurement 
basis as $\{\cos\theta\ket{00}+\sin\theta\ket{11},-\sin\theta\ket{00}+\cos\theta\ket{11},\cos\eta\ket{01}+
\sin\eta\ket{10},-\sin\eta\ket{01}+\cos\eta\ket{10}\}$, where $\theta \in [0,\pi]$ and $\eta \in [0,\pi]$.

We will consider the following three qubit states to show the usefulness of the dissension vectors. Let us consider 
a product state, 
$\rho_{pro}=\ket{000}\bra{000}$ and a classical state, $\rho_{ccc}=\frac{1}{2}(\ket{000}\bra{000}+\ket{111}\bra{111})$. 
The later has only classical correlations but former has none. Examples of three qubit separable states which may 
have only local quantumness,  
are $\rho_{ccq}=\frac{1}{2}(\ket{00+}\bra{00+}+\ket{110}\bra{110})$, 
$\rho_{qqc}=\frac{1}{2}(\ket{++0}\bra{++0}+\ket{001}\bra{001})$, and 
$\rho_{qqq}=\frac{1}{2}(\ket{+++}\bra{+++}+\ket{000}\bra{000})$. We further consider two classes of 
pure highly entangled states 
-- GHZ state, $\ket{G_3}=\frac{1}{\sqrt{2}}(\ket{000}+\ket{111})$ and W-state, 
$\ket{W}=\frac{1}{\sqrt{3}}(\ket{100}+\ket{010}+\ket{001})$.
From the Table \ref{3qtrac1}, it is clear that the dissension vector, $\vec{\delta}_1^1\simeq \kappa \vec{\delta}_1^2$, 
where $\kappa$ is 
just a scale factor. It seems that only one of them is sufficient for our characterization and following the similar reasoning, 
we find that the dissension vector, $\vec{\delta}_2^1$ captures more information than the symmetric quantity, 
$\delta_2^2$. Then we may not need the Track II dissensions at all. However, this might not be the case 
always as one will find in the following subsection.

It is evident from the Table \ref{3qtrac1} that the dissension vectors characterizes 
the correlations in the above three-qubit states. For the state, $\rho_{pro}$, 
the dissension  vectors are null, depicting it does not have any local as well as 
nonlocal quantumness. The state $\rho_{ccc}$ has only classical correlations, no local and 
nonlocal quantumness whereas the states, $\rho_{ccq}$, $\rho_{qqc}$ and $\rho_{qqq}$ have 
only local quantumness. From the Table \ref{3qtrac1}, we observe that the dissension
vectors based on the measurement on one qubit, $\vec{\delta}_1^1$ and  $\vec{\delta}_1^2$,
quantify correlations, both classical and quantum, or more accurately mixedness of qubits.
In the case of separable states, classical state qubits are maximally mixed. The vector
is most negative for this state. In the case of pure entangled states, individual qubits of 
the system in GHZ state are maximally mixed, and the dissension vector is most negative
for this state. In the case of classical state, $\rho_{ccc}$, and GHZ-state,  $\rho_{G_3}$,  
the system qubits are maximally mixed, and the dissension vectors are identical. On the 
other hand, the dissension vector based on two-qubit measurements, $\vec{\delta}_2^1$,
is positive and increases with the quantumness of the state. In the case of separable
states, from $\rho_{ccc}$ to $\rho_{qqq}$, as quantumness is increasing, this vector
is becoming larger. It takes larger values for entangled states and is largest
for GHZ-state, as one would expect.

\begin{center}
\begin{figure}[h]
\[
\begin{array}{ccc}
\includegraphics[height=7cm,width=13cm]{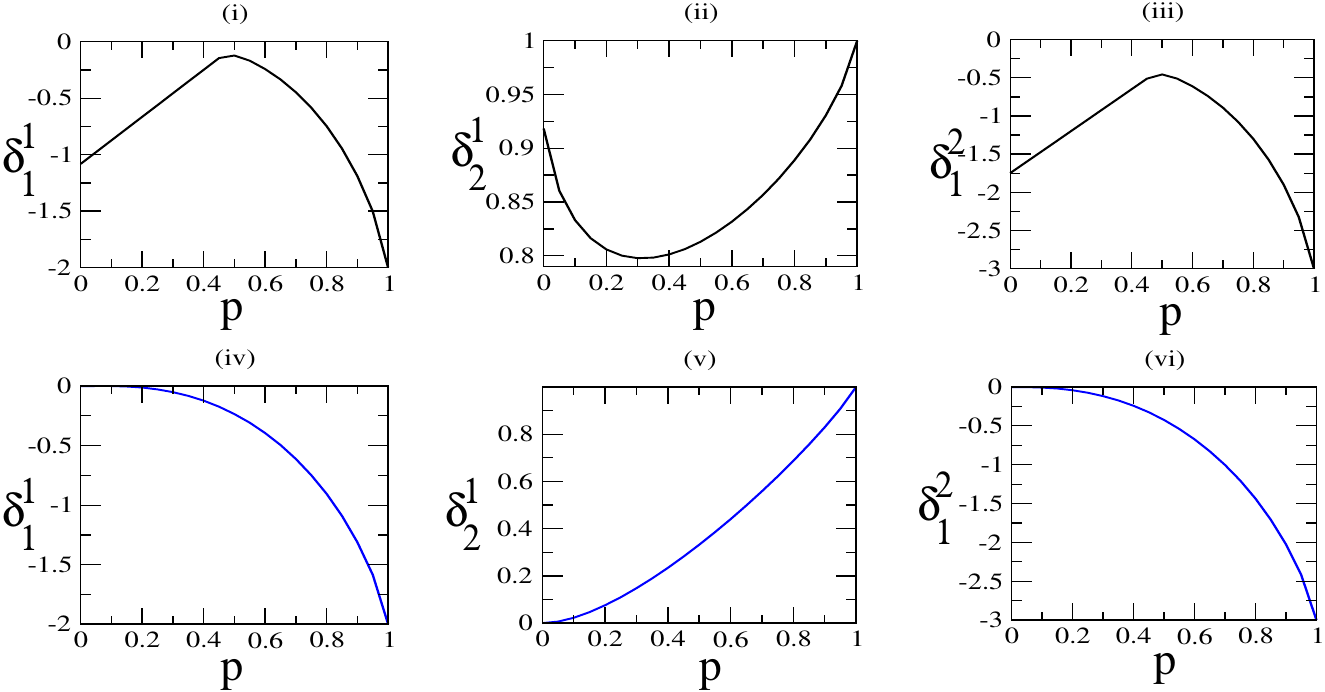}
\end{array}
\]
\caption{(Color online) The figure shows how the dissension vectors behave as a function 
of mixing parameter $p$ for the three qubit mixed states 
$\rho_{Gw}$ and $\rho_{wer}$. The subfigures [$(i)-(iii)$] 
depict the behavior of the dissensions for the state 
$\rho_{Gw}$ and [$(iv)-(vi)$] 
for $\rho_{wer}$ where subfigures $(i)$ and $(iv)$ 
depict $\vec{\delta}_1^1$; $(ii)$ and $(v)$ depict 
$\vec{\delta}_2^1$ and $(iii)$ and $(vi)$ depict $\vec{\delta}_1^2$. 
(Note that we have plotted one of the elements from each dissension vectors. 
This is because within a vector each elements are same as the states are symmetric.)}
\label{comp1}
\end{figure}
\end{center}

To illustrate the above facts, we also consider the following mixed entangled states, 
$\rho_{WG}=(1-p)\rho_{W}+p\rho_{G_3}$
and $\rho_{Wer}=\frac{(1-p)}{8} I+p\rho_{G_3}$.
Out of these 
states, $\rho_{wer}$ can be thought as the Werner state in three qubit scenario.
In Fig.(\ref{comp1}) we have plotted the behavior of the dissension vectors as a 
function of mixing parameter $p$. We note from the Fig.(\ref{comp1}), that for both
mixed states,  $\vec{\delta}_1^1$ and  $\vec{\delta}_1^2$ are more negative for
larger correlations, while  $\vec{\delta}_2^1$ is more positive with more quantumness
in the states.

\begin{table} \small
\begin{center}
\begin{tabular}{|c|c|c||c|}
\hline 
State & $\vec{\delta}_1^1$ & $\vec{\delta}_2^1$& $\vec{\delta}_1^2$\\
\hline $\rho_{pro}$& $\{0,0,0\}$ & $\{0,0,0\}$ & $\{0,0,0\}$\\
\hline $\rho_{ccc}$& $\{-2,-2,-2\}$ & $\{0,0,0\}$ & $\{-3,-3,-3\}$\\
\hline $\rho_{ccq}$& $\{-1.2,-1.2,-1.6\}$ & $\{0,0,0\}$ & $\{-1.8,-1.8,-2.6\}$\\
\hline $\rho_{qqc}$& $\{-0.99,-0.99,-0.78\}$& $\{0,0,0.22\}$ & $\{-1.6,-1.6,-1.17\}$\\
\hline $\rho_{qqq}$& $\{-0.67,-0.67,-0.67\}$& $\{0.15,0.15,0.15\}$ & $\{-1.06,-1.06,-1.06\}$\\
\hline $\rho_{G_3}$  & $\{-2,-2,-2\}$ & $\{1,1,1\}$ &  $\{-3,-3,-3\}$\\
\hline $\rho_{W}$& $\{-1.08,-1.08,-1.08\}$& $\{0.92,0.92,0.92\}$ & $\{-1.75,-1.75,-1.75\}$\\
\hline
\end{tabular}
\captionof{table}{Track-I \& II dissension vectors for a few three-qubit states. 
Note that we haven't considered $\delta_2^2$ 
because it is simply equal to the sum of elements in $\vec{\delta}_1^2$.
}
\label{3qtrac1}
\end{center}
%

\end{table}

\subsection{Four-qubit states}
In case of four-qubit system, there are three dissension vectors, ($\vec{\delta}_1^1$, $\vec{\delta}_2^1$, $\vec{\delta}_3^1$)
in track-I and two vectors, ($\vec{\delta}_1^2$, $\vec{\delta}_2^2$) and one symmetric discord, $\delta_3^2$ in track-II. The single 
qubit and two qubit measurement strategies are discussed above. For three qubit measurements we have considered eight orthogonal non-maximally entangled 
three qubit GHZ class states with at least four parameters on which we will perform optimization.

In the previous subsection, we find that the dissension vectors from one track might suffice to describe the 
quantumness of a multiparticle state. Below, we discuss that for four qubit states, some dissension vectors from 
track-II are necessary to distinguish them. Hence, we may stick with either track-II dissension vectors or consider 
more relevant dissension vectors from both the tracks. 
The second choice seems more 
useful as we know the dissension vector which contains bipartite discords, is better in characterizing 
states than the symmetric discord one. For four qubit case, we are 
considering $\vec{\delta}_3^1$ from track-I and ($\vec{\delta}_1^2$, $\vec{\delta}_2^2$) from track-II. Furthermore, we have numerically 
checked that $\vec{\delta}_1^2=\kappa_1 \vec{\delta}_1^1$ and $\vec{\delta}_2^2=\kappa_2 \vec{\delta}_2^1$, where $\kappa_i$ are scale factor. 

Let us consider following states: a product state, $\rho_{pro}=\ketbra{0000}{0000}$, a classical state, 
$\rho_{cl}=\frac{1}{2}(\ket{0000}\bra{0000}+\ket{1111}\bra{1111})$, some separable states with local quantumness, eg.,
$\rho_{3cq}=\frac{1}{2}(\ket{000+}\bra{000+}+\ket{1110}\bra{1110})$, $\rho_{2q2c}=\frac{1}{2}(\ket{++00}\bra{++00}+\ket{0011}\bra{0011})$, 
$\rho_{3qc}=\frac{1}{2}(\ket{+++0}\bra{+++0}+\ket{0001}\bra{0001})$ and $\rho_{4q}=\frac{1}{2}(\ket{++++}\bra{++++}+\ket{0000}\bra{0000})$. 
Then we consider some pure highly entangled states, eg., the GHZ state, $\ket{G_4}=\frac{1}{\sqrt{2}}(\ket{0000}+\ket{1111})$, 
the W-state, $\ket{W}=\frac{1}{2}(\ket{1000}+\ket{0100}+\ket{0010}+\ket{0001})$ and 
the $\Omega$-state, $\ket{\Omega}=\frac{1}{\sqrt{2}}(\ket{0\psi^{+}0}+\ket{1\psi^{-}1})$, 
where $\ket{\psi^{\pm}}=\frac{1}{\sqrt{2}}(\ket{00}\pm\ket{11})$. 

From Table \ref{4qtrac2}, it is clear that the state $\rho_{pro}$ has no quantumness. 
The state $\rho_{cl}$ has only classical correlations. Local quantumness is increasing
in other listed separable states ($\rho_{3cq}$, $\rho_{2c2q}$, $\rho_{c3q}$ and $\rho_{4q}$). 
The behavior of $\vec{\delta}_3^1$ is same as that of $\vec{\delta}_2^1$ for three-qubit states. It captures the quantumness of a state. It becomes larger for a more quantum state.
The behavior of  $\vec{\delta}_2^2$ is similar to  $\vec{\delta}_1^2$ in the 
three-qubit case. It is more negative for for states with more correlations, whether
classical or quantum. This suggests, that for a $n$-qubit state, dissension vectors based on
$(n-1)$-qubit measurement
quantify quantumness of a state and become more positive with increasing
quantumness. On the other hand, dissension vectors based on $(n-2)$-qubit 
measurement quantify correlation, both classical and quantum, and become
more negative with increasing correlations.

\begin{center}
\begin{figure}[h]
\[
\begin{array}{ccc}
\includegraphics[height=3.7cm,width=15cm]{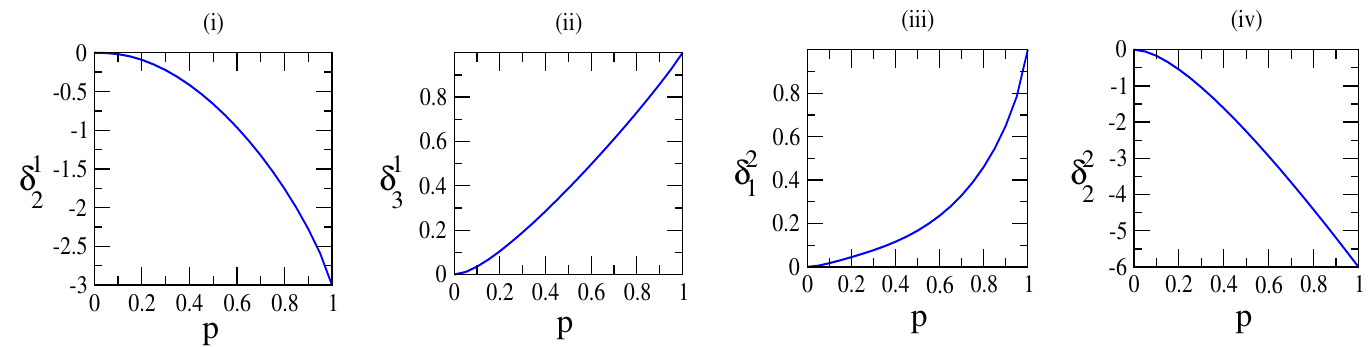}
\end{array}
\]
\caption{(Color online) The figure shows how the dissension vectors behave as a function of mixing parameter, 
$p$, for the four qubit state $\rho_{Wer}$.  
(Note that we have plotted one of the elements from each dissension vectors. 
This is because within a vector each elements are same as the states are symmetric.)}
\label{compwer}
\end{figure}
\end{center}

The four qubit mixed state, Werner-like 
state, $\rho_{Wer}=\frac{(1-p)}{16} I+p\rho_{G_4}$ show the similar features.
The dissension vectors are plotted in Fig.(\ref{compwer}). The plots indicate that as $p$ approaches 
unity (i.e., state becoming more entangled), the dissension vectors, $\vec{\delta}_2^1$ and 
$\vec{\delta}_2^2$ are becoming more negative, while $\vec{\delta}_3^1$ is becoming more
positive.

%

\begin{table} \footnotesize
\centering
\begin{tabular}{|c|c|c|c|}
\hline
$State$             & $\vec{\delta}_1^2$ & $\vec{\delta}_2^2$ & $\vec{\delta}_3^1$\\
\hline
$\rho_{pro}$        & $\{0,0,0,0\}$ & $\{0,0,0,0\}$ & $ \{0,0,0,0\} $ \\
\hline
$\rho_{cl}$       & $\{0,0,0,0\}$ & $\{-6,-6,-6,-6\}$ & $ \{0,0,0,0\} $ \\
\hline
$\rho_{3cq}$                   & $\{0.2,0.2,0.2,0.2\}$ & $\{-6,-6,-5.6,-4.8\}$ & $ \{0,0,0,0\} $ \\
\hline
$\rho_{2c2q}$                   & $\{0.35,0.29,0.35,0.35\}$ & $\{-5.6,-5.2,-4.4,-4.6\}$ & $ \{0,0,0,0\} $ \\
\hline
$\rho_{c3q}$                   & $\{0.4,0.4,0.46,0.46\}$ & $\{-4.56,-3.92,-3.92,-4.14\}$ & $ \{0.18,0,0,0\} $ \\
\hline
$\rho_{4q}$       & $\{0.5,0.5,0.5,0.5\}$ & $\{-3.35,-3.35,-3.35,-3.35\}$ & $\{0.11,0.11,0.11,0.11\} $ \\
\hline
$\rho_{W}$                   & $\{1.3,1.3,1.3,1.3\}$ & $\{-4.75,-4.75,-4.75,-4.75\}$ & $ \{0.81,0.81,0.81,0.81\} $ \\
\hline
$\rho_{G_4}$        & $ \{1,1,1,1\} $ & $\{-6,-6,-6,-6\}$ & $ \{1,1,1,1\} $ \\
\hline
$\rho_{\Omega}$                   & $\{-2,-2,-2,-2\}$ & $\{-10,-10,-10,-10\}$ & $ \{1,1,1,1\} $ \\
\hline
\end{tabular}
\captionof{table}{The dissension vectors for a few four-qubit states. Both track-I and II dissension vectors are considered here.}
\label{4qtrac2}

\end{table}

\section{Some features of dissension vectors}
Here in this section, we look into some attributes of dissension vectors.
\subsection{Why Track-II is necessary?}
Let us consider the following states, 
$\ket{\psi}_{1\shortmid 234}=\ket{0}\ket{G_3}$, $\rho_{1\shortmid 234}^c
=\frac{1}{2}(\ket{0 G_3}\bra{0 G_3}$ $+$ $\ket{1 G_3}\bra{1 G_3})$ and 
$\rho_{1\shortmid 234}^q=\frac{1}{2}(\ket{0 G_3}\bra{0 G_3}+\ket{+ G_3}\bra{+ G_3})$. These states are 
different in the sense that the first state is product 
in in $1|234$ cut but other two are not. Last one has local quantumness in $1|234$ cut. From the 
Table \ref{4qtracs1}, it is clear that only using track-I 
dissension vectors, we cannot distinguish the above states but it will be possible to distinguish them 
if we consider the dissension vector $\vec{\delta}_2^2$ 
from track-II. So, sometime the track-II dissension vectors are needed.
\begin{center}
\begin{tabular}{|c|c|c|c|c|}
\hline
$State$             & $\vec{\delta}_1^1$ ($\vec{\delta}_1^2$) & $\vec{\delta}_2^1$ & $\vec{\delta}_3^1$& $\vec{\delta}_2^2$\\
\hline
$\rho_{1\shortmid 234}$         & $ \{-1,0,0,0\} $ & $\{0,-3,-3,-2\}$ & $\{0,1,1,1\}$ & $\{-3,-6,-6,-5\}$ \\
\hline
$\rho_{1\shortmid 234}^c$         & $ \{-1,0,0,0\} $ & $\{0,-3,-3,-2\}$ & $\{0,1,1,1\}$ & $\{-6,-7,-7,-6\}$ \\
\hline
$\rho_{1\shortmid 234}^q$         & $ \{-1,0,0,0\} $ & $\{0,-3,-3,-2\}$ & $\{0,1,1,1\}$ & $\{-4.8,-6.6,-6.6,-5.6\}$ \\
\hline
\end{tabular}
\captionof{table}{Track-I and II dissension vectors for a few specific four qubit states. 
The table shows that one will not be able to 
distinguish the states from the Track-I dissension vectors whereas one will be if 
he/she considers Track-II dissension vectors. (Note that for this particular case, $\vec{\delta}_1^1\equiv\vec{\delta}_1^2$.)}
\label{4qtracs1}
\end{center} 
\subsection{Average quantumness of multiqubit states}

A vector measure characterizes a state in a fine-grained manner. Sometime,
one may be interested in average correlation properties. For some quantum
tasks, average properties may be relevant. For such tasks, two states with
different vector measures, but same `average' properties may both be suitable.
Therefore, in this section, we consider average of the dissension vectors. 
We will investigate if our measures are 
good in characterizing the states if we take average in a particular dissension 
quantity. Let us define the average dissension quantities,
\begin{equation}
	\langle\delta_1^{\ell}\rangle=\frac{1}{n}\displaystyle\sum_{k=1}^n\delta_{1k}^{\ell},
\end{equation}
where $\ell=1,2$ denotes the track in which we are calculating them. Similarly, 
we can have different quantities like $\{\langle\delta_i^{\ell}\rangle;i=1,2,...,n-1\}$, 
except the quantity, $\delta_{n-1}^{2}$ which is a symmetric quantity and sum of 
all bipartite discord. Here, we will illustrate these measures particularly for 
some three qubit states. 
 
\begin{table}[h]
\begin{centering}
\begin{tabular}{|c|c|c||c|}
\hline
State & $\langle\delta_1^1\rangle$ & $\langle\delta_2^1\rangle$& $\langle\delta_1^2\rangle$ \\ 
\hline $\ket{\psi}_{i\shortmid jk}$  & $-\frac{1}{3}$ & $\frac{2}{3}$ & $-\frac{2}{3}$\\
\hline $\rho_{cc\overleftarrow{q}}$& $-\frac{4}{3}$ & $0$& $-\frac{31}{15}$\\
\hline $\rho_{qq\overleftarrow{c}}$& $- 0.92$& $0.07$& $- 1.46$\\
\hline
\end{tabular}
\caption{Track-I average dissensions for few three-qubit states: 
$\ket{\psi}_{i\shortmid jk}=\ket{0}_i\ket{G_2}_{jk}$, 
$\rho_{ccq}$ and $\rho_{qqc}$. Here $\leftarrow$ on $q/c$ indicates that moving them towards left, 
we find other states which have same average dissension.}\label{3_av-diss1}
\end{centering}
\end{table}
Results are presented in the Table \ref{3_av-diss1}. As expected, once we look at the average 
properties, some states cannot be distinguished. For example $\rho_{ccq}$, $\rho_{cqc}$, and $\rho_{qcc}$ have 
same average quantumness but have different entries in the dissension vector $\vec{\delta}_1^{2}$ i.e., $\{-1.8,-1.8,-2.6\}$, 
$\{-1.8,-2.6,-1.8\}$ and $\{-2.6,-1.8,-1.8\}$ respectively. (Same problem arises for the vector $\vec{\delta}_1^{1}$.) 
Similar analysis goes for the states, ($\ket{\psi}_{1\shortmid 23}$, $\ket{\psi}_{2\shortmid 13}$, 
$\ket{\psi}_{3\shortmid 12}$) and ($\rho_{qqc}$, $\rho_{qcq}$, $\rho_{cqq}$). Therefore, it will be always 
advantageous if we consider dissension vectors instead of their averages.

\subsection{Behavior of quantumness under local noise}

For almost all quantum processing devices, effect of noise is inevitable. This leads us to examine
the behavior of our dissension vector under local noise. 
From a property of a measure of quantum correlations, e.g. $Q$,
for the bipartite state $\rho_{12}$,
\begin{equation}
 Q(\rho_{12})\geq Q(\Lambda_{12}[\rho_{12}]),
\end{equation}
where $\Lambda_{12}=\Lambda_1\otimes\Lambda_2$ are local channels. Under global operations, 
the situation may be different. One can create or increase entanglement under such
operations.  

\begin{center}
\begin{figure}[h]
\[
\begin{array}{ccc}
\includegraphics[height=3.5cm,width=12cm]{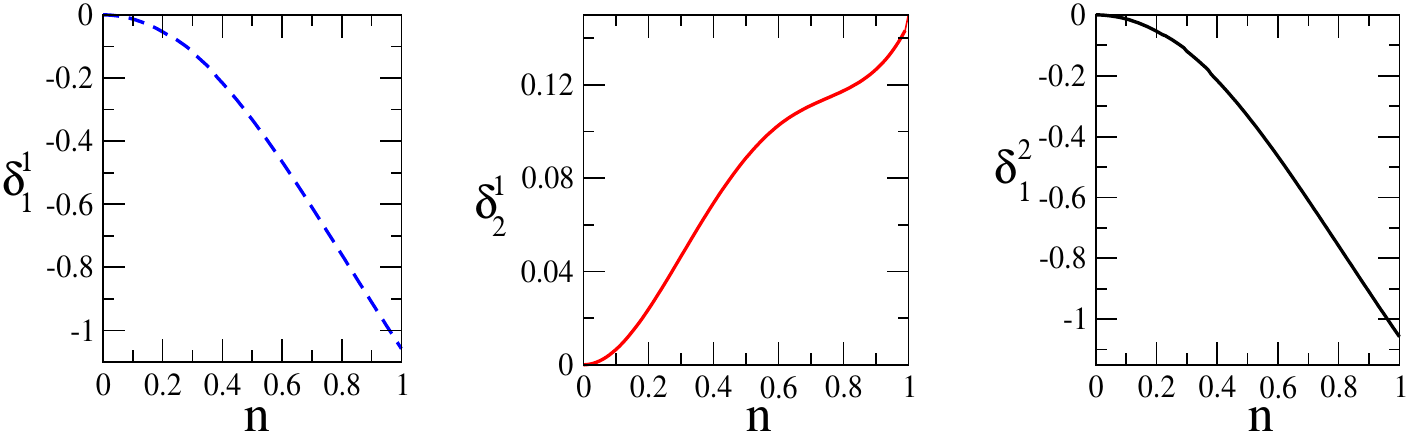}
\end{array}
\]
\caption{(Color online) The figure depicts the behavior of the ``dissension vectors'' of $\rho_{cl}$ under 
non-unital channel parameter $n$. Here, the non-unital channel ($\Lambda_{nu}$) is applied 
on the first qubit. The dissension vectors, $\vec{\delta}^1_1$ (blue dashed line), $\vec{\delta}^1_2$ 
(red solid line), and $\vec{\delta}^2_1$ (black solid line) are nonzero for finite value of $n$. 
(Note that we have plotted one of the elements from each dissension vectors. 
This is because within a vector each elements are same.)}
\label{compome1}
\end{figure}
\end{center}

It is evident that our measures are also affected by the local noise. In this 
respect we can define two important classes of channels- a unital/semiclassical channel 
$\Lambda_{u/sc}$ is defined as $\Lambda_{u/sc}(\frac{\mathbb{I}}{2})=\frac{\mathbb{I}}{2}$ 
while for a non-unital channel $\Lambda_{nu}$, $\Lambda_{nu}(\frac{\mathbb{I}}{2})\neq\frac{\mathbb{I}}{2}$. 
Streltsov et al. \cite{PhysRevLett.107.170502} have shown that a local quantum channel acting on a 
single qubit can create `quantumness' in a multiqubit system iff it is neither 
semiclassical nor unital. This result holds for the dissension vector also.
In our vector type measure, at least one of the elements will be affected. 
For example, let us consider a classical state 
$\rho_{cl}=\frac{1}{2}(\ket{0000}\bra{0000}+\ket{1111}\bra{1111})$. 
Now, application of non-unital channel \{$E_1=\ketbra{0}{0}$, $E_2=\ketbra{n}{1}$\} 
with $\ket{n}=\frac{1}{1+n^2}(\ket{0}+n\ket{1})$ ($n \in \mathbb{R}$) on 
any subsystem will make the state non-classical and will have 
non-zero element in the vector (see Fig.(\ref{compome1})).  
\section{Physical applications of dissension vectors}
The state of a $n$-bit classical computer can always be expressed by $n$-qubit classically correlated states. 
This is because a classically correlated states are described by joint probability distribution, and have specific form of dissension vectors. Also, there exist states other than 
classically correlated states. (Here, we are excluding product states and maximally mixed states.) Will the use of 
such states yield some quantum enhancement in the computation? Below we will show some examples where dissension 
vectors may have a role to play.
\subsection{BB84 protocol}
In 1984, Bennett and Brassard developed a cryptographic protocol \cite{QKD_BB84} where they utilized the 
quantum no-cloning theorem \cite{Nature_NClone_82}. In this protocol Alice prepares a qubit in one of the four states $\{\ket{0},\ket{1},\ket{+},\ket{-}\}$, and 
sends it to Bob. Bob measures the state in either $\{\ket{0},\ket{1}\}$ basis or in $\{\ket{+},\ket{-}\}$ randomly. This 
process goes for many rounds and at the end both publicly declare the basis of preparation and measurement respectively. 
There are two possible outcomes -- 1. basis match i.e., perfect correlations 
and 2. basis do not match. If in second case they find any correlation then they may conclude that someone (Eve) was 
eavesdropping in between the state transmission. Otherwise they will use the the correlated bit string (of case 1) 
as one-time pad, the fundamental object of cryptography.

In the whole process the average state of Alice and Bob is 
\begin{equation}
 \rho_{BB84}=\frac{1}{4}(\ketbra{00}{00}+\ketbra{11}{11}+\ketbra{0+}{0+}+\ketbra{1-}{1-}).
\end{equation}
The state $\rho_{BB84}$ is a mixture of two maximally classically correlated states in two different basis. The state has 
no entanglement. The state has non-zero local quantumness in Bob's part. This means local quantumness on Bob's side 
is providing the security in the protocol. Let us examine it. 

Assume that to read the key, Eve will perform a universal state independent Buzek-Hillary (B-H) cloning 
operations\cite{BHclone} on the state while it in transmission from 
Alice to Bob. Now, before applying the cloning operation the average initial state of Alice, Bob and Eve is 
$\rho_i=\rho_{BB84}\otimes \ketbra{0}{0}_{eve}$. After the Eve's action, the state becomes 
\begin{equation}
 \rho_f=\frac{1}{4}\left[\ketbra{000}{000}+\ketbra{111}{111}+\frac{2}{3}\mathbb{I}_2\otimes\ketbra{\phi^+}{\phi^+}
 +\frac{1}{3}\{\ketbra{011}{011}+\ketbra{100}{100}+\sigma_z\otimes(\ketbra{\psi^+}{\phi^+}
 +\ketbra{\phi^+}{\psi^+})\}\right],
\end{equation}
where $\ket{\psi^+}=\frac{1}{\sqrt{2}}(\ket{00}+\ket{11})$, $\ket{\phi^+}=\frac{1}{\sqrt{2}}(\ket{01}+\ket{10})$, 
$\sigma_z=\ketbra{0}{0}-\ketbra{1}{1}$, and $\mathbb{I}_2$ is a identity matrix of order $2$. The dissension vectors for 
the states $\rho_i$ and $\rho_f$ are are given in Table \ref{BB84}.
\begin{table}[h]
\begin{centering}
\begin{tabular}{|c|c|c||c|}
\hline
State & $\vec{\delta_1^1}$ & $\vec{\delta_2^1}$ & $\vec{\delta_1^2}$\\ 
\hline $\rho_i$& $\{0,0,-0.40\}$& $\{0,0,0\}$ & $\{0,0,-0.79\}$\\
\hline $\rho_f$& $\{-0.40,-0.24,-0.24\}$& $\{0,0.29,0.29\}$& $\{-0.73,-0.41,-0.41\}$\\
\hline
\end{tabular}
\caption{Dissensions for the states in the BB84 protocol.}\label{BB84}
\end{centering}
\end{table}
From dissension vectors, it is evident that the correlations in the state $\rho_f$ has increased due to 
Eve's action. And hence, the loss of security of BB84 protocol may be decided based on dissension vectors.
\subsection{Dissension in Grover search algorithm}
Grover search algorithm was introduced for accelerating the data search from an `unstructured database' 
\cite{PhysRevLett.79.325, RevModPhys.82.1}. 
It was believed that entanglement may be necessary to achieve such speed up \cite{Pro.R.Soc.Lond.A.495}. Later, it was shown 
that the entanglement is not directly related to the probability of success in the search \cite{QIC.2.399}. 
Also, it is not clear whether there is a relationship between the probability of success and the correlations that go
beyond entanglement (particularly captured by well known measure, quantum discord) \cite{1751-8121-43-4-045305}. 
Recently, it was suggested that the success probability relies on the depletion of quantum coherence 
\cite{2016arXiv161008656S, 2016arXiv161104542A}.

Here, we investigate possible role of the dissension vectors in Grover search algorithm. In 
this algorithm, the initial $n$-qubit database can be expressed by 
\begin{equation}
 \ket{\psi_0}=\sqrt{\frac{j}{2^n}}\ket{\chi}+\sqrt{1-\frac{j}{2^n}}\ket{\chi^{\perp}},
\end{equation}
where $j$ are the number of solutions in Grover search algorithm and $\ket{\chi}=\frac{1}{\sqrt{j}}\sum_{x}\ket{x}$ 
($\{\ket{\chi},\ket{\chi^{\perp}}\}$ form a basis). In the next step, a Grover operation (called iteration) is 
applied repeatedly to improve the proportion of solutions. The Grover operation, $G=\mathbb{A}\mathcal{O}$, consists of 
Oracle, $\mathcal{O}=\mathbb{I}-2\braket{\chi}{\chi}$ and an inversion operation, 
$\mathbb{A}=2\braket{\psi_0}{\psi_0}-\mathbb{I}$. After $r$ iterations, the global state takes the form
\begin{equation}
 \ket{\psi_r}=G^r\ket{\psi_0}=\sin\theta_r\ket{\chi}+\cos\theta_r\ket{\chi^{\perp}},
\end{equation}
where $\theta_r=(r+\frac{1}{2})\beta$ and $\beta=2 \arctan \sqrt{\frac{j}{2^n-j}}$. The final step is to 
obtain $\ket{\chi}$ with high probability by performing measurement on $\ket{\psi_r}$. The probability of the 
success is given by $P_{succ}=\sin^2\theta_r$. The optimal time to stop the iteration is 
at $r_{opt}=C_I[\frac{\pi-\beta}{2\beta}]$ times, where $C_I[m]$ denotes the closest integer to $m$.

\begin{center}
\begin{figure}[h]
\[
\begin{array}{ccc}
\includegraphics[height=8cm,width=11cm]{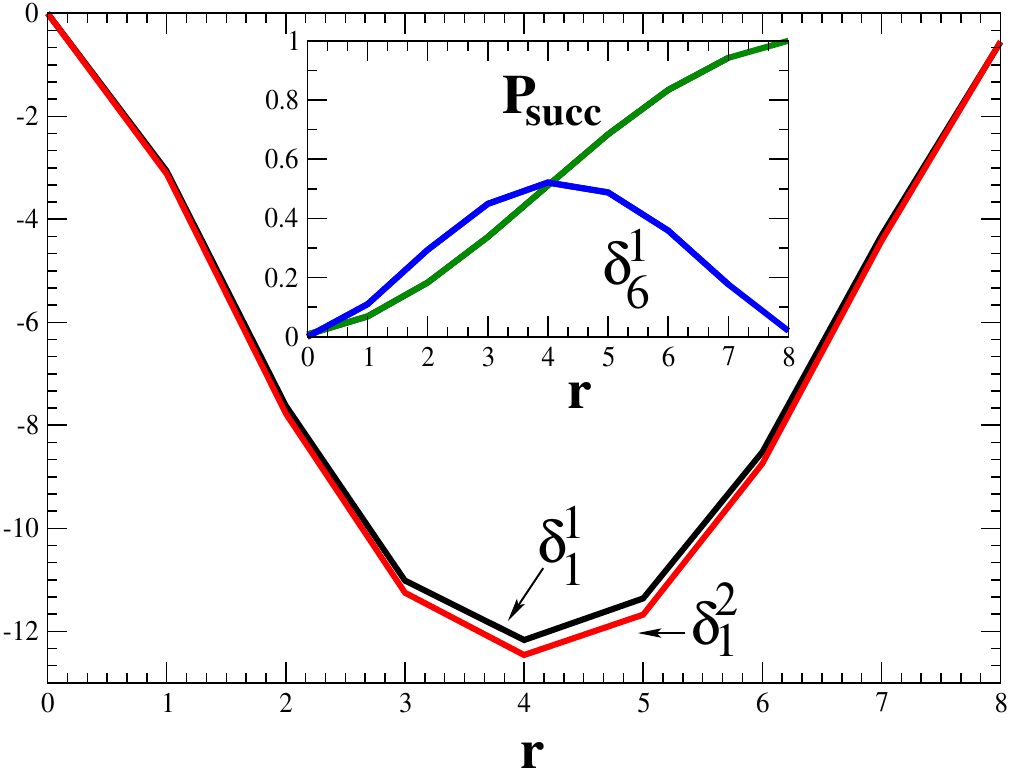}
\end{array}
\]
\caption{(Color online) The figure depicts the behavior of the dissension vectors of $\rho_{r}$ with the 
Grover iteration $r$. The inset figure shows the plot of success probability with $r$. 
It indicates that while the dissension vectors $\vec{\delta}_1^1$ and $\vec{\delta}_1^2$ are decreasing with $r$ and reaches its 
minimum at $r=4$, the $\vec{\delta}_6^1$ is increasing upto $r=4$ but remains always positive. It shows that the 
correlations during the Grover search increases with $r$ and reaches its maximum then starts decreasing.}
\label{groverf}
\end{figure}
\end{center}

We will only consider the simplest situation of single solution (i.e., $j=1$) and will assume that the 
solution is located at $\ket{0}$. Then the final density matrix generated by Grover search has the following form 
\begin{equation}\label{rho-1}
\rho_r=\left(
  \begin{array}{ccccc}
    a^2 & ab  & ab  & ab  & \cdots \\
    ab  & b^2 & b^2 & b^2 & \cdots \\
    ab  & b^2 & b^2 & b^2 & \cdots \\
    ab  & b^2 & b^2 & b^2 & \cdots \\
    \vdots & \vdots & \vdots & \vdots & \ddots \\
  \end{array}
\right)
_{2^n\times 2^n},
\end{equation}
where $a=\sin\theta_r$ and $b=\frac{1}{\sqrt{2^n-j}}\cos\theta_r$. 
And the reduced density matrix of any $k$-qubits is defined as 

\begin{equation}\label{rho-reduced}
\rho_r^k=\left(
  \begin{array}{ccccc}
    a^2+(2^{n-k}-1)b^2 & ab+(2^{n-k}-1)b^2  & ab+(2^{n-k}-1)b^2  & ab+(2^{n-k}-1)b^2  & \cdots \\
    ab+(2^{n-k}-1)b^2  & 2^{n-k}b^2         & 2^{n-k}b^2         & 2^{n-k}b^2         & \cdots \\
    ab+(2^{n-k}-1)b^2  & 2^{n-k}b^2         & 2^{n-k}b^2         & 2^{n-k}b^2         & \cdots \\
    ab+(2^{n-k}-1)b^2  & 2^{n-k}b^2         & 2^{n-k}b^2         & 2^{n-k}b^2         & \cdots \\
    \vdots & \vdots & \vdots & \vdots & \ddots \\
  \end{array}
\right)
_{2^k\times 2^k}.
\end{equation}
We investigate behavior of the dissension vectors of the state $\rho_r$ for $n=7$. The quantum 
correlations in the state increases with the Grover iterations and reaches its maximum at some $r$ value (here it is 
$4$). Then it starts to decrease and becomes zero when the final projective measurement is performed 
to obtain the solution (see Fig.[\ref{groverf}]).

Recently, in Ref.\cite{2014arXiv1410.7067C}, it was shown that the some modified form of dissensions may be useful 
in characterizing 
the average state merging cost. Apart from that the dissensions may be useful in multiparticle entanglement distributions 
\cite{PhysRevLett.108.250501,PhysRevLett.109.070501}, 
quantum cryptography \cite{srep06956} and quantum interferometry \cite{PhysRevLett.112.210401}.

\section{Conclusion}

By considering the dissension vector as a measure of the quantumness of a 
multiqubit state, we have argued that a vector quantity, as a fine-grained measure,
does a better job in characterizing and quantifying the quantum properties of a state. 
We considered two tracks of these measures for $n$-qubit states. In particular,
for three-qubit and four-qubit systems, we showed how various classes of states 
can be distinguished and characterized using these measures. In particular,
we saw that in the case of $n$-qubit states, for ($n-2$)-qubit measurements,
the dissension vectors $\vec{\delta}_{n-2}^1$ and $\vec{\delta}_{n-2}^2$ quantify
correlations, both classical and quantum. More correlated states have more 
negative values for these vectors. On the other hand, for ($n-1$)-qubit measurements,
the dissension vector $\vec{\delta}_{n-1}^1$ quantifies quantumness of a state and is
always positive. We have also considered the effect of local noise, and how to
quantify average quantumness.
We also discussed applications of these measures in the context of
BB84 protocol and Grover search algorithm.

%

{\it Acknowledgement}: Author SS would like to thank Mr. Abhishek Deshpande, 
Dr. Indranil Chakraborty and Prof. V. Ravishankar for having useful discussions. We thank anonymous referee for 
his valuable comments.



\section*{References}
\begin{thebibliography}{0}

\bibitem{RevModPhys.84.1655}
Kavan Modi, Aharon Brodutch, Hugo Cable, Tomasz Paterek, and Vlatko Vedral.
\newblock The classical-quantum boundary for correlations: Discord and related
  measures.
\newblock {\em Rev. Mod. Phys.}, 84:1655--1707, Nov 2012.

\bibitem{RevModPhys.81.865}
Ryszard Horodecki, Pawe\l{} Horodecki, Micha\l{} Horodecki, and Karol
  Horodecki.
\newblock Quantum entanglement.
\newblock {\em Rev. Mod. Phys.}, 81:865--942, Jun 2009.

\bibitem{2005quant.ph..8124J}
R.~{Jozsa}.
\newblock {An introduction to measurement based quantum computation}.
\newblock {\em eprint arXiv:quant-ph/0508124}, August 2005, quant-ph/0508124.

\bibitem{PhysRevA.84.022324}
Rafael Chaves and Fernando de~Melo.
\newblock Noisy one-way quantum computations: The role of correlations.
\newblock {\em Phys. Rev. A}, 84:022324, Aug 2011.

\bibitem{PhysRevLett.100.050502}
 Datta, A. Shaji, A. and Caves, C. M. 
\newblock Quantum discord and the power of one qubit.
\newblock {\em Phys. Rev. Lett.}, 100:050502, Feb 2008.

\bibitem{RevModPhys.74.145}
Nicolas Gisin, Gr\'egoire Ribordy, Wolfgang Tittel, and Hugo Zbinden.
\newblock Quantum cryptography.
\newblock {\em Rev. Mod. Phys.}, 74:145--195, Mar 2002.

\bibitem{PhysRevLett.96.010401}
V. Giovannetti, S. Lloyd, and  L. Maccone, 

\newblock {\em Phys. Rev. Lett.}, 96:010401, Jan 2006.

\bibitem{PhysRevLett.88.017901}
Harold Ollivier and Wojciech~H. Zurek.
\newblock Quantum discord: A measure of the quantumness of correlations.
\newblock {\em Phys. Rev. Lett.}, 88:017901, Dec 2001.

\bibitem{discordHend2001}
L~Henderson and V~Vedral.
\newblock Classical, quantum and total correlations.
\newblock {\em Journal of Physics A: Mathematical and General}, 34(35):6899,
  2001.

\bibitem{PhysRevA.71.062307}
Micha\l{} Horodecki, Pawe\l{} Horodecki, Ryszard Horodecki, Jonathan Oppenheim,
  Aditi Sen(De), Ujjwal Sen, and Barbara Synak-Radtke.
\newblock Local versus nonlocal information in quantum-information theory:
  Formalism and phenomena.
\newblock {\em Phys. Rev. A}, 71:062307, Jun 2005.

\bibitem{PhysRevLett.100.140502}
A.~R.~Usha Devi and A.~K. Rajagopal.
\newblock Generalized information theoretic measure to discern the quantumness
  of correlations.
\newblock {\em Phys. Rev. Lett.}, 100:140502, Apr 2008.

\bibitem{PhysRevLett.104.080501}
Kavan Modi, Tomasz Paterek, Wonmin Son, Vlatko Vedral, and Mark Williamson.
\newblock Unified view of quantum and classical correlations.
\newblock {\em Phys. Rev. Lett.}, 104:080501, Feb 2010.

\bibitem{okrasa2011}
M.~Okrasa and Z.~Walczak.
\newblock Quantum discord and multipartite correlations.
\newblock {\em EPL (Europhysics Letters)}, 96(6):60003, 2011.

\bibitem{PhysRevA.84.042109}
C.~C. Rulli and M.~S. Sarandy.
\newblock Global quantum discord in multipartite systems.
\newblock {\em Phys. Rev. A}, 84:042109, Oct 2011.

\bibitem{Chakrabarty2011}
I.~Chakrabarty, P.~Agrawal, and A.~K. Pati.
\newblock Quantum dissension: Generalizing quantum discord for three-qubit
  states.
\newblock {\em The European Physical Journal D}, 65(3):605--612, 2011.

\bibitem{lafla2002}
Raymond Laflamme, D.~Cory, C.~Negrevergne, and L.~Viola.
\newblock {NMR} quantum information processing and entanglement.
\newblock {\em Quantum Inf. Comput.}, 2:166 -- 176, Feb 2002.

\bibitem{PhysRevA.83.032323}
Vaibhav Madhok and Animesh Datta.
\newblock Interpreting quantum discord through quantum state merging.
\newblock {\em Phys. Rev. A}, 83:032323, Mar 2011.

\bibitem{PhysRevLett.90.100402}
Micha\l{} Horodecki, Karol Horodecki, Pawe\l{} Horodecki, Ryszard Horodecki,
  Jonathan Oppenheim, Aditi Sen(De), and Ujjwal Sen.
\newblock Local information as a resource in distributed quantum systems.
\newblock {\em Phys. Rev. Lett.}, 90:100402, Mar 2003.

\bibitem{PhysRevLett.89.180402}
Jonathan Oppenheim, Micha\l{} Horodecki, Pawe\l{} Horodecki, and Ryszard
  Horodecki.
\newblock Thermodynamical approach to quantifying quantum correlations.
\newblock {\em Phys. Rev. Lett.}, 89:180402, Oct 2002.

\bibitem{PhysRevA.66.022104}
A.~K. Rajagopal and R.~W. Rendell.
\newblock Separability and correlations in composite states based on entropy
  methods.
\newblock {\em Phys. Rev. A}, 66:022104, Aug 2002.

\bibitem{PhysRevLett.105.190502}
Borivoje Daki\ifmmode~\acute{c}\else \'{c}\fi{}, Vlatko Vedral, and \ifmmode
  \check{C}\else~\v{C}\fi{}aslav Brukner.
\newblock Necessary and sufficient condition for nonzero quantum discord.
\newblock {\em Phys. Rev. Lett.}, 105:190502, Nov 2010.

\bibitem{2015arXiv150200857A}
P.~{Agrawal}, I.~{Chakrabarty}, S.~{Sazim}, and A.~K. {Pati}.
\newblock {Local, nonlocal quantumness and information theoretic measures}.
\newblock {\em  Int. J. Quantum Inform.}, 14:1640034 Sept 2016, arXiv:1502.00857.

\bibitem{PhysRevA.93.062322}
G.~Bellomo, A.~Plastino, and A.~R. Plastino.
\newblock Quantumness and the role of locality on quantum correlations.
\newblock {\em Phys. Rev. A}, 93:062322, Jun 2016.

\bibitem{PhysRevLett.107.170502}
Alexander Streltsov, Hermann Kampermann, and Dagmar Bru\ss{}.
\newblock Behavior of quantum correlations under local noise.
\newblock {\em Phys. Rev. Lett.}, 107:170502, Oct 2011.

\bibitem{PhysRevA.88.022315}
Gian~Luca Giorgi.
\newblock Quantum discord and remote state preparation.
\newblock {\em Phys. Rev. A}, 88:022315, Aug 2013.

\bibitem{PhysRevLett.112.140507}
Pawe\l{} Horodecki, Jan Tuziemski, Pawe\l{} Mazurek, and Ryszard Horodecki.
\newblock Can communication power of separable correlations exceed that of
  entanglement resource?
\newblock {\em Phys. Rev. Lett.}, 112:140507, Apr 2014.

\bibitem{PhysRevLett.107.190501}
Gian~Luca Giorgi, Bruno Bellomo, Fernando Galve, and Roberta Zambrini.
\newblock Genuine quantum and classical correlations in multipartite systems.
\newblock {\em Phys. Rev. Lett.}, 107:190501, Nov 2011.

\bibitem{PhysRevA.72.032317}
Berry Groisman, Sandu Popescu, and Andreas Winter.
\newblock Quantum, classical, and total amount of correlations in a quantum
  state.
\newblock {\em Phys. Rev. A}, 72:032317, Sep 2005.

\bibitem{2009PhLA..373.1818W}
Z.~{Walczak}.
\newblock {Total correlations and mutual information}.
\newblock {\em Physics Letters A}, 373:1818--1822, May 2009, arXiv: 0806.4861.

\bibitem{PhysRevA.76.032327}
Nan Li and Shunlong Luo.
\newblock Total versus quantum correlations in quantum states.
\newblock {\em Phys. Rev. A}, 76:032327, Sep 2007.

\bibitem{PhysRevLett.79.5194}
N.~J. Cerf and C.~Adami.
\newblock Negative entropy and information in quantum mechanics.
\newblock {\em Phys. Rev. Lett.}, 79:5194--5197, Dec 1997.

\bibitem{PhysRevA.60.893}
N.~J. Cerf and C.~Adami.
\newblock Quantum extension of conditional probability.
\newblock {\em Phys. Rev. A}, 60:893--897, Aug 1999.

\bibitem{2005Natur.436..673H}
M.~{Horodecki}, J.~{Oppenheim}, and A.~{Winter}.
\newblock {Partial quantum information}.
\newblock {\em Nature}, 436:673--676, August 2005, quant-ph/0505062.

\bibitem{nature.delN}
L.~{del Rio}, J.~{Aberg}, R.~{Renner}, O.~{Dahlsten}, and V.~{Vedral}.
\newblock {The thermodynamic meaning of negative entropy}.
\newblock {\em Nature}, 474:61 -- 63, june 2011.

\bibitem{Horodecki2007}
Micha{\l} Horodecki, Jonathan Oppenheim, and Andreas Winter.
\newblock Quantum state merging and negative information.
\newblock {\em Communications in Mathematical Physics}, 269(1):107--136, 2007.

\bibitem{cover1991}
T.~M. Cover and J.~A. Thomas.
\newblock {\em Elements of information theory}.
\newblock John Wiley and Sons, Inc., New York, USA, 1991.

\bibitem{MI-Herbut2004}
F.~Herbut.
\newblock On mutual information in multipartite quantum states and equality in
  strong subadditivity of entropy.
\newblock {\em Journal of Physics A: Mathematical and General}, 37(10):3535,
  2004.

\bibitem{PhysRevA.66.042309}
N.~J. Cerf, S.~Massar, and S.~Schneider.
\newblock Multipartite classical and quantum secrecy monotones.
\newblock {\em Phys. Rev. A}, 66:042309, Oct 2002.

\bibitem{2015arXiv150407176K}
A.~{Kumar}.
\newblock {Multiparty quantum mutual information: An alternative definition}.
\newblock {\em Phys. Rev. A}, 96: 012332, July 2017.

\bibitem{JHEP145} 
Dawei Ding, Patrick Hayden, and Michael Walter.  
\newblock {Conditional mutual information of bipartite unitaries and scrambling}.
\newblock {\em Journal of High Energy Physics}, 12:145, December 2016, arXiv:1608.04750.

\bibitem{2017arXiv170302903S} 
{Sharma}, K. and {Wakakuwa}, E. and {Wilde}, M.~M.
\newblock{Conditional quantum one-time pad}.
\newblock {\em ArXiv e-prints}, March 2017, 1703.02903.

\bibitem{QKD_BB84}

\newblock {\em Quantum Cryptography: Public key distribution and coin tossing.} 
\newblock Los Alamitos, CA, 1984.

\bibitem{Nature_NClone_82}
W.~K. Wootters and W.~H. Zurek.
\newblock {A single quantum cannot be cloned.}
\newblock {\em Nature}, 299:802 -- 803, Oct 1982.

\bibitem{BHclone}

\newblock {The universal state independent B-H cloning operations 
\cite{PhysRevA.54.1844, PhysRevLett.81.5003} are defined as
\begin{eqnarray}
 \ket{0}\ket{0}\rightarrow \sqrt{\frac{2}{3}}(\ket{00}+\ket{\phi^+}),\nonumber\\
 \ket{1}\ket{0}\rightarrow \sqrt{\frac{2}{3}}(\ket{11}+\ket{\phi^+}),\nonumber
\end{eqnarray}
where first ket in the left hand side is the target and second one is the ancilla where the target ket will be 
copied}.

\bibitem{PhysRevLett.79.325}
Grover, Lov K.
\newblock {Quantum Mechanics Helps in Searching for a Needle in a Haystack}.
\newblock {\em Phys. Rev. Lett.}, 79:325--328, Jul 1997.

\bibitem{RevModPhys.82.1}
Childs, Andrew M. and van Dam, Wi.
\newblock {Quantum algorithms for algebraic problems}.
\newblock {\em Rev. Mod. Phys.}, 82:1--52 Jan 2010.

\bibitem{Pro.R.Soc.Lond.A.495}
Richard Jozsa, Noah Linden. 
\newblock {On the role of entanglement in quantum-computational speed-up}. 
\newblock {\em Proc. R. Soc. Lond. A}, 459:2011, 2003.

\bibitem{QIC.2.399}
Braunstein S.~L, Pati, A.~K.
\newblock {Speed-up and entanglement in quantum searching}. 
\newblock {\em Quantum Inform. Computat.}, 2(5): 399--409, 2002.

\bibitem{1751-8121-43-4-045305}
Cui, Jian and Fan, Heng.
\newblock {Correlations in the Grover search}". 
\newblock {\em Journal of Physics A: Mathematical and Theoretical}, 43:045305, 2010.

\bibitem{2016arXiv161008656S}
  {Shi}, H.-L. and {Liu}, S.-Y. and {Wang}, X.-H. and {Yang}, W.-L. and {Yang}, Z.-Y. and {Fan}, H.
\newblock {Coherence depletion in the Grover quantum search algorithm}.
\newblock {\em ArXiv e-prints}, Oct 2016, 1610.08656.

\bibitem{2016arXiv161104542A}
{Anand}, N. and {Pati}, A.~K.
\newblock {Coherence and Entanglement Monogamy in the Discrete Analogue of Analog Grover Search}.
\newblock {\em ArXiv e-prints}, Nov 2016, 1611.04542.

\bibitem{2014arXiv1410.7067C}
I. {Chakrabarty}, and {Deshpande}, A. and {Chatterjee}, S.
\newblock{Quantum Residual Correlation: Interpreting through State Merging}.
\newblock {\em ArXiv e-prints}, Oct 2014, 1410.7067.

\bibitem{PhysRevLett.108.250501}
Streltsov, Alexander and Kampermann, Hermann and Bru\ss{}, Dagmar.
\newblock{Quantum Cost for Sending Entanglement}.
\newblock {\em Phys. Rev. Lett.} 108:250501, Jun 2012.

\bibitem{PhysRevLett.109.070501}
Chuan, T. K. and Maillard, J. and Modi, K. and Paterek, T. and Paternostro, M. and Piani, M.
\newblock{Quantum Discord Bounds the Amount of Distributed Entanglement}.
\newblock {\em Phys. Rev. Lett.}, 109:070501, Aug 2012.

\bibitem{PhysRevLett.112.210401}
Girolami, Davide and Souza, Alexandre M. and Giovannetti, Vittorio and Tufarelli, Tommaso and 
Filgueiras, Jefferson G. and Sarthour, Roberto S. and Soares-Pinto, Diogo O. and Oliveira, 
Ivan S. and Adesso, Gerardo.
\newblock{Quantum Discord Determines the Interferometric Power of Quantum States}.
\newblock {\em Phys. Rev. Lett.}, 112:210401, May 2014.

\bibitem{srep06956}
Pirandola Stefano.
\newblock{Quantum discord as a resource for quantum cryptography}.
\newblock {\em Scientific Reports}, 4:6956, Nov 2014.



\bibitem{PhysRevA.54.1844}
Bu\ifmmode \check{z}\else \v{z}\fi{}ek, V. and Hillery, M.
\newblock {Quantum copying: Beyond the no-cloning theorem}.
\newblock {\em Phys. Rev. A}, 54:1844--1852, Sep 1996.


\bibitem{PhysRevLett.81.5003}
Bu\ifmmode \check{z}\else \v{z}\fi{}ek, Vladim\'{\i}r and Hillery, Mark.
\newblock {Universal Optimal Cloning of Arbitrary Quantum States: From Qubits to Quantum Registers}.
\newblock {\em Phys. Rev. Lett.}, 81:5003--5006, Nov 1998.
  

\end{thebibliography}
\end{document}